\documentclass[prd,twocolumn,aps,nofootinbib,nobibnotes,preprintnumbers,superscriptaddress]{revtex4}
\usepackage[dvipdfmx]{graphicx}
\usepackage{bmpsize}
\usepackage{mathrsfs}
\usepackage{bm,color,braket}
\usepackage{xcolor}
\usepackage{amsmath}
\usepackage{amssymb}
\graphicspath{{./fig/}{./png/}}

\newcommand{\BoldVec}[1]{\mathchoice%
  {\mbox{\boldmath $\displaystyle     #1$}}%
  {\mbox{\boldmath $\textstyle        #1$}}%
  {\mbox{\boldmath $\scriptstyle      #1$}}%
  {\mbox{\boldmath $\scriptscriptstyle#1$}}%
}
\newcommand{\EQ}{\begin{equation}}
\newcommand{\EN}{\end{equation}}
\newcommand{\EQA}{\begin{eqnarray}}
\newcommand{\ENA}{\end{eqnarray}}

\newcommand{\Eq}[1]{Eq.~(\ref{#1})}
\newcommand{\Eqs}[2]{Eqs.~(\ref{#1}) and~(\ref{#2})}

\newcommand{\Eqss}[2]{Eqs.~(\ref{#1})--(\ref{#2})}

\newcommand{\bbra}[1]{\left\langle #1\right\rangle}

{}
{}
{}

\newcommand{\rad}{{\rm rad}}
\newcommand{\tot}{{\rm tot}}
\newcommand{\sat}{{\rm sat}}
\newcommand{\phys}{{\rm phys}}
\newcommand{\xx}{\BoldVec{x}{}}

\newcommand{\uu}{\BoldVec{u} {}}

\newcommand{\BB}{\BoldVec{B} {}}

\newcommand{\AAA}{\BoldVec{A} {}}

\newcommand{\jj}{\BoldVec{j} {}}
\newcommand{\JJ}{\BoldVec{J} {}}

\newcommand{\EE}{\BoldVec{E} {}}

\newcommand{\kk}{\BoldVec{k} {}}

\newcommand{\nab}{\BoldVec{\nabla} {}}

\newcommand{\SSSS}{\bm{\mathsf{S}}}

\newcommand{\ii}{{\rm i}}

\newcommand{\dd}{{\rm d} {}}

\def\half{{\textstyle{1\over2}}}

\newcommand{\MeV}{\,{\rm MeV}}
\newcommand{\GeV}{\,{\rm GeV}}
\newcommand{\TeV}{\,{\rm TeV}}

\newcommand{\G}{\,{\rm G}}

\newcommand{\kg}{\,{\rm kg}}

\newcommand{\Mpc}{\,{\rm Mpc}}

\newcommand{\Mpl}{M_\mathrm{pl}}

\newcommand{\CPI}[1]{{\text{\sc cpi}}}
\newcommand{\CME}[1]{{\text{\sc cme}}}
\newcommand{\cMHD}{\texorpdfstring{$\chi$MHD}{chiMHD}}

\newcommand{\IGMF}[1]{\text{$\mathrm{IGMF}$}}
\newcommand{\erm}{{\mathrm{e}}}

\usepackage[normalem]{ulem} 

\usepackage{nicefrac}
\usepackage[hidelinks]{hyperref}
\hypersetup{
    colorlinks=true,       
    linkcolor=red,          
    citecolor=blue,        
    filecolor=magenta,      
    urlcolor=blue           
}
\usepackage[all]{hypcap} 

\newcommand{\fref}[1]{Fig.~\ref{#1}}
\newcommand{\Fref}[1]{Figure~\ref{#1}}
\newcommand{\tref}[1]{Table~\ref{#1}}
\newcommand{\rref}[1]{Ref.~\cite{#1}}

\begin{document}

\title{Primordial magnetic field from chiral plasma instability with sourcing}

\date{\today}
\preprint{NORDITA-2025-146}

\author{Murman~Gurgenidze}
\email{mgurgeni@andrew.cmu.edu}
\affiliation{McWilliams Center for Cosmology and Department of Physics, Carnegie Mellon University, Pittsburgh, PA 15213, USA}
\affiliation{School of Natural Sciences and Medicine, Ilia State University, 0194 Tbilisi, Georgia}

\author{Andrew~J.~Long}
\email{andrewjlong@rice.edu}
\affiliation{Department of Physics and Astronomy, Rice University, 6100 Main St., Houston, TX 77005, USA}

\author{Alberto~Roper~Pol}
\email{alberto.roperpol@unige.ch}
\affiliation{D\'epartement de Physique Th\'eorique, Universit\'e de Gen\'eve, CH-1211 Gen\'eve,
Switzerland}

\author{Axel~Brandenburg}
\email{brandenb@nordita.org}
\affiliation{McWilliams Center for Cosmology and Department of Physics, Carnegie Mellon University, Pittsburgh, PA 15213, USA}
\affiliation{School of Natural Sciences and Medicine, Ilia State University, 0194 Tbilisi, Georgia}
\affiliation{Nordita, KTH Royal Institute of Technology and Stockholm University, 10691 Stockholm, Sweden}
\affiliation{Department of Astronomy, AlbaNova University Center, Stockholm University, 10691 Stockholm, Sweden} 

\author{Tina~Kahniashvili}
\email{tinatin@andrew.cmu.edu}
\affiliation{McWilliams Center for Cosmology and Department of Physics, Carnegie Mellon University, Pittsburgh, PA 15213, USA}
\affiliation{School of Natural Sciences and Medicine, Ilia State University, 0194 Tbilisi, Georgia}
\affiliation{Department of Theoretical Astrophysics and Cosmology, Georgian National Astrophysical Observatory, Tbilisi, 47/57 M. Kostava St., GE-0179, Georgia}

\begin{abstract}
In an electron-positron plasma, an imbalance in the number of right- and left-chiral particles can lead to the growth of a helical magnetic field through a phenomenon called the chiral plasma instability (CPI).  In the early universe, scattering reactions that violate chirality come into thermal equilibrium when the plasma cools below a temperature of approximately $80 \, \mathrm{TeV}$.  Since these reactions tend to relax any pre-existing chiral asymmetry to zero as the system approaches equilibrium, the standard lore is that primordial magnetogenesis via the CPI is not viable below $80 \, \mathrm{TeV}$.  In this work, we propose that the presence of a source for chirality can allow the CPI to operate even below $80 \, \mathrm{TeV}$, we explore the implications of this scenario, and we derive predictions for the resultant magnetic field helicity using a combination of analytical methods and direct numerical simulation.
\end{abstract}

\maketitle

\section{Introduction}
\label{sec:intro}

There is a rich interplay between particle asymmetries and primordial magnetism in the early universe.  
Particle asymmetries are expected to arise in the early universe during the epoch of baryogenesis \cite{Sakharov:1967dj,Pilaftsis:1997jf,Shaposhnikov:1987tw} at which time the cosmological excess of matter over antimatter was established.  
Prior work has shown that a chiral asymmetry, corresponding to an excess (or deficit) of right-chiral fermions over left-chiral fermions, leads to an electric current that can produce magnetic fields, known as the chiral magnetic effect (CME) \cite{Vilenkin:1980fu}.
First proposed in \rref{Joyce:1997uy}, various studies have explored the development of an instability leading to the growth of helical magnetic fields, known as the chiral plasma instability (CPI) \cite{Joyce:1997uy,Boyarsky:2011uy,Boyarsky:2012ex,Tashiro:2012mf,Akamatsu:2013pjd,Boyarsky:2015faa,Sigl:2015xva,Long:2016uez,Pavlovic:2016gac,Rogachevskii:2017uyc,Brandenburg:2017rcb,Schober:2017cdw,Pavlovic:2018jiz,Schober:2021iws,Schober:2023zxl}, using both analytical methods and direct numerical simulation of the nonlinear equations of motion of chiral magnetohydrodynamics (\cMHD{})  \cite{Giovannini:2013oga,Boyarsky:2015faa,Yamamoto:2016xtu,Hattori:2017usa,Brandenburg:2017rcb,Schober:2017cdw}.
See also the review articles \cite{Fukushima:2008xe,Kharzeev:2013ffa,Liao:2014ava,Kharzeev:2015znc,Kamada:2022nyt,Kharzeev:2023zbo,Li:2025yxx}.

The primordial magnetic field that is generated in this way is expected to survive in the universe today as a relic of the early universe.
In the voids between galaxy clusters, the primordial magnetic field would constitute an intergalactic magnetic field, which can be probed with astrophysical observations \cite{Neronov:2010gir}.
This field may have played an essential role in seeding the galactic dynamo by raising the effective seed magnetic field above $\sim 10^{-20} \, \mathrm{G}$, thereby helping to explain the origin of $\mu\mathrm{G}$ galactic fields through dynamo theory \cite{Durrer:2013pga,Jedamzik:2018itu, Subramanian:2019jyd,Vachaspati:2020blt}.
Furthermore, the presence of a primordial magnetic field during recombination could have led to a reduction of the sound horizon due to baryon clumping \cite{Jedamzik:2011cu,Jedamzik:2018itu}.
This effect would modify the cosmological estimate of the present-time Hubble rate from CMB data, potentially alleviating the Hubble tension \cite{Jedamzik:2020krr,Galli:2021mxk,Mirpoorian:2024fka,Jedamzik:2025cax}. 
Since the origin of cosmological magnetism remains an open question, it makes sense to ask whether such fields may have arisen in the early universe as a byproduct of baryogenesis. 

While particle asymmetries associated with some charges, such as the baryon number minus the lepton number ($\textsf{B} - \textsf{L}$), are predicted to be exactly conserved, the particle asymmetry associated with chirality is not conserved in the Standard Model.  
Rather, scatterings of chiral fermions can change their chirality if the scattering is mediated by the Yukawa interaction with the Higgs boson (or if the fermion's Dirac mass is involved). 
Once these scatterings come into thermal equilibrium in the primordial plasma, they deplete the chiral asymmetry to zero exponentially quickly (assuming that the asymmetry is not also being sourced).
Studies of these scattering processes \cite{Campbell:1990fa,Joyce:1997uy,Bodeker:2019ajh} reveal that the chirality should be depleted to zero when the primordial plasma cools to a temperature of approximately $k_B T \approx 80 \TeV$.  
At this time, the age of the Universe was only about $4 \times 10^{-17} \sec$.
This means that the CPI must take place above $80 \TeV$, before the chiral asymmetry is erased.  

However, the preceding discussion is modified if the chiral asymmetry is being sourced.  
For instance, baryogenesis generically entails an out-of-equilibrium scattering or decay that sources a chiral asymmetry and leads to the baryon asymmetry of the universe.  
In this work, we study the development of the CPI in the presence of a source for chirality in order to derive predictions for the relic primordial magnetic field.  
We argue that the source term allows the CPI to be active even at temperatures below the nominal chiral erasure temperature of $80 \TeV$.  
In some studies, the chiral charge erasure was neglected and the CPI was considered at temperatures below $80 \TeV$ \cite{Boyarsky:2011uy,Rogachevskii:2017uyc,Brandenburg:2017rcb,Schober:2018wlo,Schober:2020ogz,Brandenburg:2021aln,Schober:2021yav,Schober:2024vtv}, such as during the electroweak phase transition at $100 \GeV$ or during the QCD phase transition at $100 \MeV$. 
Our work is partly motivated by the desire to justify how the CPI could occur below the nominal charge erasure temperature of $80 \TeV$.  
In many models, baryogenesis takes place at the electroweak scale or below~ \cite{Akhmedov:1998qx,Trodden:1998ym,Cline:2006ts,Elor:2018twp,Nelson:2019fln,Klaric:2021cpi}, and our work indicates that baryogenesis may be accompanied by magnetogenesis as a consequence of the chirality sourcing.

\section{A source for chirality}
\label{sec:source}

While baryogenesis is taking place in the early universe, chirality may be sourced by out-of-equilibrium scatterings, decays, or oscillations.  
Even though chirality violation is not a necessary ingredient for baryogenesis, both charge and parity violation are necessary ingredients \cite{Sakharov:1967dj}.
Given the close connection between chirality and parity, many models of baryogenesis also entail some level of chirality violation.  
In this section, we provide a simple particle physics toy model that leads to a nonzero chiral source $S_5(t)$, and we use it to motivate the functional form of $S_5(t)$ that we use for our numerical studies, which are presented in the following sections. 

Consider a metastable particle species $\phi$ that decays to chiral electron-positron pairs via two reactions: 
\begin{equation}\begin{split}
	\Gamma_R & = \Gamma(\phi \to e_R^- + e_R^+ + X) \;,\\ 
	\Gamma_L & = \Gamma(\phi \to e_L^- + e_L^+ + \bar{X}) 
	\;.
\end{split}\end{equation}
Here $e_L^-$ denotes a left-chiral electron, $e_R^-$ denotes a right-chiral electron, $e_L^+$ denotes a left-chiral positron, and $e_R^+$ denotes a left-chiral positron.  
There may be other particles in the final states, which we denote by $X$ and $\bar{X}$, and these particles do not carry any chirality.  
The reaction with rate $\Gamma_R$ increases the electron chirality by two units, while the reaction with rate $\Gamma_L$ decreases it by two units.  
Without loss of generality we can write $\Gamma_R = (1 + \epsilon) \, (\beta / 2) \, \Gamma_\phi$ and $\Gamma_L = (1 - \epsilon) \, (\beta / 2) \, \Gamma_\phi$, where $\Gamma_\phi$ is the total decay rate of a $\phi$ particle, which may include additional channels beyond the two shown above, where $\beta$ is the branching ratio into either of the chirality-changing channels, and where $\epsilon$ is the dimensionless parameter that controls chirality violation.  
Note that $0 < \beta \leq 1$ and $-1 \leq \epsilon \leq 1$, but typically $|\epsilon| \ll 1$.

We suppose that there is a population of $\phi$ particles having an approximately homogeneous number density $n_\phi(t_c)$ at time\footnote{
In the flat Friedman-Lemaitre-Robertson-Walker cosmology, the invariant interval is $(\dd s)^2 = - (\dd t_c)^2 + a^2 (t_c) |\dd \xx|^2 = - a^2(t) (\dd t)^2 + a^2 (t) |\dd \xx|^2$, where $a$ is the scale factor, $t_c$ is the cosmic time, $t$ is the conformal time, and $\dd t = \dd t_c/a$.
} $t_c$. 
We assume that these particles decay out of equilibrium, which means that the inverse processes, including $e_R^- + X \to e_L^- + \phi$, are effectively shut off.  
If $\epsilon = 0$ such that $\Gamma_R = \Gamma_L$, then on average no chirality is generated when the population of $\phi$ particles decays. 
In contrast, if $\epsilon \neq 0$, then the decaying population of $\phi$ particles is a source of chirality.  
Let $n_5(\xx,t_c)$ represent the density of chirality at position $\xx$ and time $t_c$, i.e., the number of right-chiral fermions minus the number of left-chiral fermions per unit volume.  
For a system at temperature $T$ with chiral chemical potential $\mu_5 = (\mu_R - \mu_L) / 2 = (\mu_{e_R^-} + \mu_{e_R^+} - \mu_{e_L^-} - \mu_{e_L^+}) / 4$ this density is given by $n_5 = (k_B^2 / [\hbar^3 c^3]) (\mu_5 T^2 / 3)$.  
The rate of chirality generation per unit volume is given by 
\begin{align}\label{eq:dn5dt}
    \tfrac{\partial}{\partial t_c} n_5(\xx,t_c)
    \, \supset S_5(t_c)
	\;,
\end{align}
where there may be additional terms accounting for the washout or diffusion of chirality.  
In our phenomenological model, the chirality source term is given by 
\begin{align}
	S_5(t_c) 
    = \bigl( \Gamma_R - \Gamma_L \bigr) \, n_\phi(t_c) 
    = \epsilon \beta \Gamma_\phi n_\phi(t_c) 
    \;.
\end{align}
The time dependence of $S_5(t_c)$ follows the time dependence of $n_\phi(t_c)$, which we now proceed to calculate. 

We can say that the $\phi$ particles decay at a time $t_c = \tau_\phi$, where $\tau_\phi = \Gamma_\phi^{-1}$ is their lifetime.  
We assume that the $\phi$ particles are non-relativistic when they decay, allowing us to approximate their energy density by only their rest energy $\rho_\phi \approx m_\phi c^2 n_\phi$, where $m_\phi$ is the mass of a $\phi$ particle. 
When the $\phi$ particles decay, we assume that the cosmological energy budget of the universe is dominated by radiation, having temperature $T$ and energy density $\rho_\mathrm{rad} = (k_B^4 / [\hbar^3 c^3]) (\pi^2/30) g_E T^4$, where $g_E$ is the effective number of relativistic species.  
For example, the Standard Model particles contribute $g_E = 106.75$ at $k_B T \gtrsim 160 \GeV$ prior to the electroweak phase transition, and they contribute $g_E = 61.75$ at $k_B T \approx 1 \GeV$ prior to the QCD phase transition.
The $\phi$ particles should compose a sub-dominant component of the energy budget, and therefore it is useful to define $\Omega_\phi = \rho_\phi(\tau_\phi) / \rho_\mathrm{rad}(\tau_\phi)$, which gives their fraction of the energy budget at the time when they decay; note that $0 < \Omega_\phi < 1$.  
We can use this relation to exchange the variable $n_\phi(\tau_\phi)$ for the variable $\Omega_\phi$.  

\begin{figure}[t]
	\centering
	\includegraphics[width=0.95\linewidth]{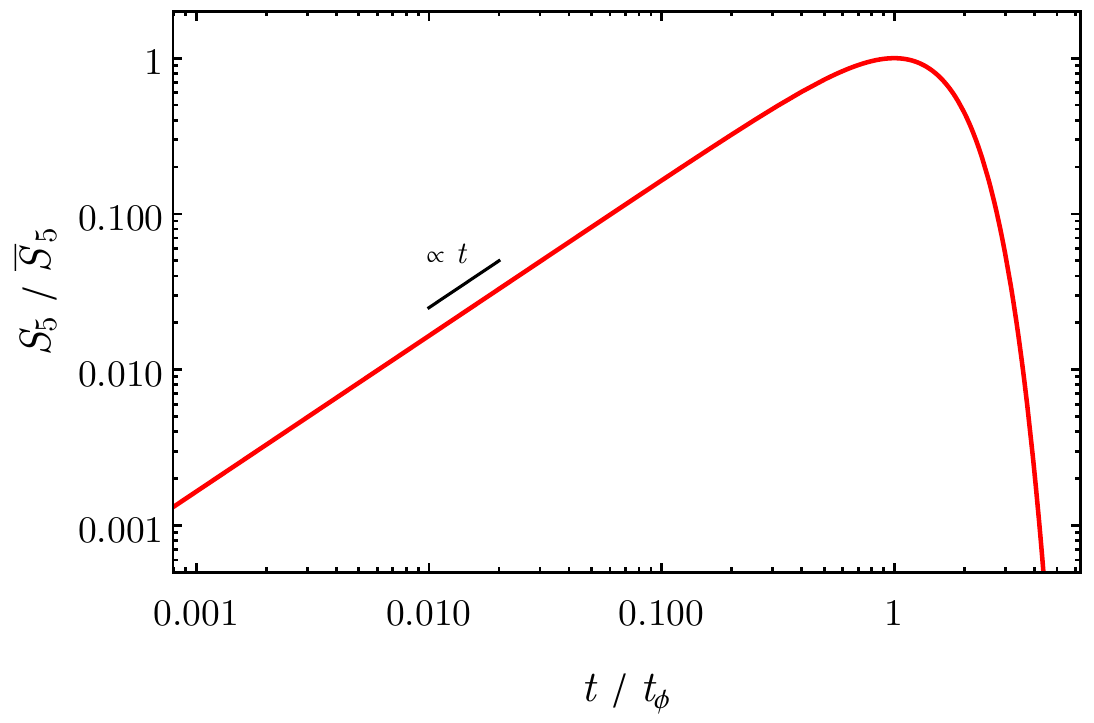} \hfill
	\caption{\label{fig:S5}
	Evolution of the chiral source $S_5(t)$ with conformal time $t$. The maximum occurs at $t = t_\phi$ where $S_5(t_\phi) = \bar{S}_5$.   
	}
\end{figure}

As a result of the cosmological expansion, the number density of massive $\phi$ particles will decrease $\propto a^{-3}(t_c)$.  
In a radiation-dominated universe, the evolution of the scale factor with cosmic time is $a(t_c) = a(\tau_\phi) \, (t_c / \tau_\phi)^{1/2}$.  
As a result of the decays, the density will decrease further $\propto \mathrm{exp}(-\Gamma_\phi t_c)$.  
Putting together these various factors, we can write the chirality source term as follows,  
\begin{align}\label{eq:S5_original}
    S_5(t_c) 
    = \epsilon \beta \Gamma_\phi \frac{\rho_\mathrm{rad}(\tau_\phi) \Omega_\phi}{m_\phi c^2}  &\,  \biggl( \frac{t_c}{\tau_\phi} \biggr)^{\!\!-3/2} 
    \\ \nonumber \times \, 
    \mathrm{exp}[-&\, \Gamma_\phi (t_c-\tau_\phi)] 
    \;.
\end{align}
Notice that $S_5(t_c)$ diverges as $t_c \to 0$, due to the $a^{-3}(t_c)$ factor.  
However, this divergence is removed if we work instead with comoving densities and conformal time.  
The comoving density obeys an equation that takes the same form as \Eq{eq:dn5dt} but where the term on the right side is instead multiplied by a factor of $a^4 (t_c)$.  
We also exchange cosmic time $t_c$ for conformal time $t$ using $t_c = a^2 (\tau_\phi)\, t^2 / (4 \tau_\phi)$, since $a \sim t_c^{1/2} \sim t$ in the radiation-dominated era.
After these changes, the comoving source term is 
\begin{align}\label{eq:S5}
    S_5(t) = \bar{S}_5 \, \frac{t}{t_\phi} \, \erm^{- (t^2 - t_\phi^2) / (2 t_\phi^2)} 
    \;,
\end{align}
where $S_5$ has been scaled with $a^4$, and the parameters $\bar{S}_5$ and $t_\phi$ are given by 
\begin{align}\label{eq:barS5}
	\bar{S}_5 = \epsilon \beta \sqrt{\frac{\erm}{2}} \frac{\rho_\mathrm{rad} (\tau_\phi) \Omega_\phi}{m_\phi c^2} \frac{a^4 (\tau_\phi)}{\tau_\phi} 
	\quad \text{and} \quad 
	t_\phi = \frac{\sqrt{2} \tau_\phi}{a(\tau_\phi)} 
    \;.
\end{align}
Here we have used $\Gamma_\phi = \tau_\phi^{-1}$ and $\erm \approx 2.718$ is the base of the natural logarithm.  
We present a graph of $S_5(t)$ in \fref{fig:S5}. 
Note that $S_5(t)$ increases linearly in the early stages, reaches a maximum of $\bar{S}_5$ when $t = t_\phi$, and falls exponentially in the later stages.  
We use this parameterization for our numerical studies.
Note that $\sqrt{2} t_\phi$ is the value of the conformal time $t$ when the cosmic time is $t_c = \tau_\phi$ in the radiation era.

\section{Equations of \cMHD{}}
\label{sec:cMHD}

The dynamical variables in \cMHD{} are the magnetic vector potential $\AAA(\xx,t)$, the magnetic field $\BB(\xx,t)$, the electric scalar potential $V(\xx,t)$, the electric field $\EE(\xx,t)$, the electric charge density $Q(\xx,t)$, the electric current density $\JJ(\xx,t)$, the fluid energy density $\rho(\xx,t)$, the fluid pressure $p(\xx,t)$, the fluid velocity $\uu(\xx,t)$, and the chiral chemical potential $\mu_5(\xx,t)$.  
The fields are related by $\EE = - \nab V - \tfrac{1}{c} \tfrac{\partial}{\partial t} \AAA$ and $\BB = \nab \times \AAA$, and we work in the Weyl gauge $V(\xx,t) = 0$. 
We assume:  i) a relativistic equation of state, which restricts $p = \rho/3$; ii) an electrically neutral plasma, which restricts $Q = 0$; and iii) a constitutive relation for the current, which restricts $\JJ = \JJ_\mathrm{Ohm} + \JJ_\CME{}$, where $\JJ_\mathrm{Ohm} = \sigma (\EE + \tfrac{1}{c} \uu \times \BB)$ is the Ohmic current, $\sigma$ is the electric conductivity, $\JJ_\CME{} = c \tilde{\mu}_5 \BB$ is the chiral magnetic effect current~\cite{Vilenkin:1980fu}, $\tilde{\mu}_5(\xx,t) = 2 \alpha \mu_5(\xx,t) / (\pi \hbar c)$ is the rescaled chiral chemical potential with units of $\mathrm{[length]}^{-1}$, and $\alpha = e^2/(4\pi \hbar c) \approx 1/137$ is the electromagnetic fine structure constant. 

The equations of motion for these dynamical variables in an expanding Universe are a coupled system of partial differential equations \cite{Brandenburg:1996fc,Brandenburg:2017rcb,Rogachevskii:2017uyc,RoperPol:2025lgc} 
\begin{subequations}\label{eq:cMHD}
\begin{align}
    \label{eq:dAdt} 
    \tfrac{\partial}{\partial t} \AAA 
    & = \uu \times \BB 
    + \eta \bigl( \tilde{\mu}_5 \BB - \tfrac{1}{c} \JJ \bigr) 
    \;, 
    \\
    \label{eq:dmudt}
    \bigl( \tfrac{\partial}{\partial t} + \uu \cdot \nab \bigr) \tilde{\mu}_5  
    & = - \tilde{\mu}_5 \nab \cdot \uu  
    + D_5 \nabla^2 \tilde{\mu}_5 
    \\ & \quad 
    - \lambda \eta \bigl( \tilde{\mu}_5 \BB - \tfrac{1}{c} \JJ \bigr) \cdot \BB 
    - \Gamma_5 \tilde{\mu}_5 
    + \tilde{S}_5 
    \;,
    \nonumber \\ 
    \label{eq:dudt}
    \rho \bigl( \tfrac{\partial}{\partial t} + \uu \cdot \nab \bigr) \uu 
    & = 
    2 \nab \cdot \bigl( \rho \nu \SSSS \bigr) 
    - \tfrac{c^2}{4} \nab \rho 
    \\
    & \quad 
    + \tfrac{1}{3} \uu \nab \cdot (\rho \uu)
    + \tfrac{3c}{4} \JJ \times \BB 
    \nonumber \\ 
    & \quad 
    - \uu \bigl[ \tfrac{1}{c} \uu \cdot (\JJ \times \BB) + \tfrac{1}{c^2} \eta |\JJ|^2 \bigr] 
    \;,
    \nonumber \\
    \label{eq:dlnrhodt}
    \tfrac{\partial}{\partial t} \rho 
    & = 
    - \tfrac{4}{3} \nab \cdot (\rho \uu)
    \\ & \quad 
    + \tfrac{1}{c} \uu \cdot (\JJ\times\BB) 
    + \tfrac{1}{c^2} \eta |\JJ|^2
    \;, 
    \nonumber 
\end{align}
\end{subequations}
where ${\sf S}_{ij} = ( \partial_j u_i + \partial_i u_j ) / 2 - \delta_{ij} \partial_k u_k/3$ is the traceless rate-of-strain tensor.  
Amp\`ere's law with Maxwell's correction implies $\JJ = c \nab \times \BB - \tfrac{\partial}{\partial t} \EE$, and we neglect $\tfrac{\partial}{\partial t} \EE$.
Notice that this system of equations is valid for the subrelativistic limit\footnote{
We also note that the $\nab \cdot (\rho \uu)$ term in \Eqs{eq:dudt}{eq:dlnrhodt} has been corrected in \rref{RoperPol:2025lgc} to $\rho \nab \cdot \uu + {1 \over 2} \uu \cdot \nab \rho$ in the subrelativistic limit.  However, this modification is expected to be negligible for the CPI as velocity fields driven by the magnetic field are mostly vortical, reducing the impact of density perturbations in the dynamics.
} 
of fluid bulk motion $|\uu|^2 \ll 1$, see \rref{RoperPol:2025lgc} for the extension to relativistic $|\uu|^2$.  
The model parameters are the electric conductivity $\sigma$, the magnetic diffusivity $\eta = c^2 / \sigma$, the kinematic viscosity $\nu$, the chiral diffusion coefficient $D_5$, the chiral feedback parameter $\lambda$, the chiral erasure rate $\Gamma_5$, and the rescaled chiral source $\tilde{S}_5(t)$, which we define in the following; see \Eq{eq:S5tilde}.
Note that all of the parameters and dynamical variables are comoving quantities (i.e., they have been scaled by a number of $a(t)$ factors equal to their mass dimension); $\xx$ is the comoving spatial coordinate, $t$ is the conformal time coordinate, and $\uu$ is the peculiar velocity.
The physical variables are related to the comoving ones as $\AAA = a \AAA_{\rm phys}$, $\BB = a^2 \BB_\phys$, $\EE = a^2 \EE_\phys$, $\JJ = a^3 \JJ_\phys$, $\eta = \eta_\phys/a$, $\tilde \mu_5 = a \tilde \mu_{5, \phys}$, $\lambda = \lambda_\phys/a^2$, $\Gamma_5 = a \Gamma_{5, \phys}$, $\tilde S_5 = a^4 \tilde S_{5, \phys}$, $\rho = a^4 \rho_\phys$, $D_5 = D_{5,\phys}/a$, and $\nu = \nu_\phys/a$.

The equation of motion for the chiral chemical potential \eqref{eq:dmudt} is derived as follows. 
Let $j_5^\mu(x) = \bar{\psi} \gamma^\mu \gamma_5 \psi = \bar{\psi}_R \gamma^\mu \psi_R - \bar{\psi}_L \gamma^\mu \psi_L = (c n_5, \jj_5)$ denote the Noether current density associated with the axial-vector symmetry transformation.  
It is normalized such that the Noether charge density $n_5$ counts $+1$ for right-chiral fermions and $-1$ for left-chiral fermions.
The Noether current is not conserved locally, because the axial-vector symmetry is anomalous in QED, but also because the interactions giving rise to the washout and source terms must explicitly break the symmetry.  
Consequently, the continuity equation for chiral charge is
\begin{align}\label{eq:continuity_eqn}
    \tfrac{\partial}{\partial t} n_5 + \nab \cdot \jj_5 = 8 \tfrac{1}{\hbar} \tfrac{\alpha}{4\pi} \EE \cdot \BB - \Gamma_5 n_5 + S_5 
    \;.
\end{align}
The first term on the right is the axial-vector anomaly, $\partial_\mu j_5^\mu = - 2 \tfrac{1}{\hbar} \tfrac{\alpha}{4\pi} F_{\mu\nu} \tilde{F}^{\mu\nu}$, where the factor of $2$ accounts for electrons and positrons.
See Ch.~19 of \rref{Peskin:1995} for additional discussion.  

To evaluate the chiral current density, we assume a linear constitutive relation $\jj_5 = n_5 \uu - D_5 \nab n_5$.  
Here we have introduced the parameter $D_5$, which is the diffusion coefficient for chiral charge. 
To evaluate the chiral charge density, we assume that the system consists of an electron-positron plasma at temperature $T$ with a small chiral asymmetry ($\mu_5 \ll T$).  
Then, the chiral charge density is approximately $n_5 \approx (k_B^2 / [\hbar^3 c^3]) (\mu_5 T^2 / 3)$, where $\mu_5 = (\mu_R - \mu_L)/2 = (\mu_{e_R^+} + \mu_{e_R^-} - \mu_{e_L^+} - \mu_{e_L^-})/4$.  
Multiplying \Eq{eq:continuity_eqn} by $(\hbar^3 c^3 / k_B^2) (3 / T^2) (2 \alpha / [\pi \hbar c])$ yields \Eq{eq:dmudt}, and we identify the scaled source: 
\begin{align}\label{eq:S5tilde}
	\tilde{S}_5(t) 
    & = \bigl( \tfrac{\hbar^3 c^3}{k_B^2} \bigr) \bigl( \tfrac{3}{T^2} \bigr) \bigl( \tfrac{2 \alpha}{\pi \hbar c} \bigr) S_5(t) \\ 
	& = \tilde{\bar{S}}_5 \, \frac{t}{t_\phi} \, \erm^{- (t^2 - t_\phi^2) / (2 t_\phi^2)} 
    \;, \nonumber \\ 
    \text{where} \quad 
    \tilde{\bar S}_5 & = 3 \sqrt{g_E} \sqrt{\frac{\erm}{5}} \frac{ \Mpl \alpha}
    {c m_\phi t_\phi^2} \epsilon \beta \Omega_\phi 
    \;.
	\nonumber
\end{align}
Note that we have introduced the reduced Planck mass $\Mpl = \sqrt{(\hbar c)/(8 \pi G)} \approx 4.3 \times 10^{-9} \kg$ and used the relation $\rho_\rad (\tau_\phi) = 3 c \Mpl^2 H^2 (\tau_\phi)/\hbar =  \Mpl k_B^2 T^2 (\tau_\phi) \pi \sqrt{g_E} H(\tau_\phi)/(\sqrt{10} \hbar^2 c)$, where $H (\tau_\phi)$ is the Hubble rate at the cosmic time $\tau_\phi$.
This relation is satisfied during radiation domination. 
We can read off the chiral feedback parameter, which is found to be $\lambda = (\hbar^3 c^3 / k_B^2) (3 / T^2) (2 \alpha / [\pi \hbar c]) (8 / [\hbar c]) (\alpha / [4\pi])$ giving 
\begin{align}\label{eq:lambda}
    \lambda = \lambda_\star \equiv \biggl( \frac{\hbar c}{k_B^2} \biggr) \biggl( \frac{12 \alpha^2}{\pi^2 T^2} \biggr) 
    \;,
\end{align}
which agrees with the value used in \rref{Brandenburg:2023aco}.  
For our numerical studies, we explore different values of $\lambda$, so we denote this theoretically-preferred ``fiducial'' value by $\lambda_\star$.

It is worth reiterating that all of the formulas here are expressed in terms of comoving quantities, such as the comoving chemical potential $\mu_5 = a \mu_{5,\mathrm{phys}}$ and the comoving temperature $T = a T_\mathrm{phys}$.  
While the universe expands and cools adiabatically, $T$ is approximately static. 

\section{Chiral plasma instability}
\label{sec:CPI}

The equations of \cMHD{} admit an instability, known as the chiral plasma instability (CPI), toward the growth of a helical magnetic field \cite{Joyce:1997uy}.  
To identify the instability, it is sufficient to set $\uu = 0$ in \Eq{eq:dAdt}, which becomes 
\begin{align}
    \tfrac{\partial}{\partial t} \AAA 
    - \eta \nabla^2 \AAA 
    - \eta \tilde{\mu}_5 \nab \times \AAA 
    = 0 
    \;.
\end{align}
Note that we have used the radiation gauge with $V = 0$ and $\nab \cdot \AAA = 0$, where the latter is a consequence of $V = 0$ and the neutrality of the plasma $Q = 0$, such that $\nab \cdot \EE = 0$.  
The vector potential may be decomposed into Fourier modes $\AAA(\xx,t) = \int \!\! \tfrac{\dd^3 \kk}{(2\pi)^3} \AAA_{\kk}(t) \, \erm^{\ii \kk \cdot \xx}$, and further decomposed onto a basis of right- and left-handed circular polarization modes $\AAA_{\kk}(t) = A_{\kk,+}(t) \hat{\bm \epsilon}_{\kk,+} + A_{\kk,-}(t) \hat{\bm \epsilon}_{\kk,-}$. 
In terms of these variables, the equation of motion becomes 
\begin{align}\label{eq:A_evolution}
    \tfrac{\partial}{\partial t} A_{\kk,\pm}
    + \bigl( \eta |\kk|^2 \mp \eta \tilde{\mu}_5 |\kk| \bigr) A_{\kk,\pm} 
    = 0 
    \;.
\end{align}
If $\tilde{\mu}_5 = 0$, then $A_{\kk,\pm} \propto \mathrm{exp}[-\eta |\kk|^2 t]$ decays at a rate controlled by the magnetic diffusivity $\eta$.  
If $\tilde{\mu}_5 > 0$, then the positive-helicity modes $A_{\kk,+}$ with $|\kk| < \tilde{\mu}_5$ are unstable.  
If $\tilde{\mu}_5 < 0$, then the negative-helicity modes $A_{\kk,-}$ with $|\kk| < -\tilde{\mu}_5$ are unstable.  
This is a tachyonic instability that only impacts long-wavelength modes.  

It is customary to study the CPI in a system without chiral sources ($\tilde{S}_5 = 0$), without chirality erasure ($\Gamma_5 = 0$), and with a nonzero initial chiral
asymmetry ($\tilde{\mu}_5(t_i) \neq 0$).
For such a system, it is useful to identify $k_\CPI{} = |\tilde{\mu}_5(t_i)|$ and note that the long-wavelength modes with $|\kk| < k_\CPI{}$ are unstable.  
The instability develops most quickly for modes with wavenumber $|\kk| = k_\CPI{} / 2$, which begin to grow exponentially (while $\tilde{\mu}_5$ is initially approximately constant) on a time scale $t_\CPI{} = 4 / \eta k_\CPI{}^2$, defined such that $A_{k_\CPI{}, \pm} \propto \erm^{t/t_\CPI{}}$, where the unstable mode is determined by the sign of $\tilde \mu_5$.

As the CPI develops, the chiral asymmetry is converted into magnetic helicity density.
This can be understood from \Eq{eq:continuity_eqn} by noting that
\begin{equation}
    \tfrac{\partial}{\partial t} h_M 
    = -2 c \bbra{\EE \cdot \BB} 
    = - \tfrac{\pi c \hbar}{\alpha} \tfrac{\lambda_\star}{\lambda} \tfrac{\partial}{\partial t} \bbra{n_5} 
    \;,
\end{equation}
where we have defined the magnetic helicity density
\begin{equation}\label{eq:helicity_definition}
    h_M = \bbra{\AAA\cdot \BB} = \frac{1}{V} \int \! \dd^3 \xx \, \AAA \cdot \BB 
    \;.
\end{equation}
Angle brackets denote volume average, and the terms corresponding to divergence of fluxes vanish in a periodic volume.
Note that we have assumed $\tilde S_5 = \Gamma_5 = 0$ for this conservation law to be satisfied, and we allow $\lambda \neq \lambda_\star$ in our numerical studies.
In terms of the chiral chemical potential, the resultant comoving helicity density is estimated as 
\begin{subequations}\label{eq:Deltah}
\begin{align}
    \Delta h_M
    & = - \frac{k_B^2}{\hbar^2 c^2} \frac{\lambda_\star}{\lambda} \frac{\pi}{3 \alpha} \, \Delta \! \bbra{\mu_5} \, T^2 
    \;,
\end{align}
and recall that $\tilde{\mu}_5 = 2 \alpha \mu_5 / (\pi \hbar c)$. 
Since the derivation has assumed a small chiral asymmetry, this formula is only valid for $|\Delta \! \bbra{\mu_5}| \ll 
k_B T$.  
We estimate the maximum comoving helicity density to be 
\begin{equation}\label{eq:delta_hM}
\begin{split}
    \Delta h_M & \approx -(1.06 \times 10^{-18} \, \mathrm{G})^2 \, 
    \mathrm{Mpc} \, a_0^3 
    \\ & \qquad \times 
    \bigl( \tfrac{\lambda_\star}{\lambda} \bigr) \bigl( \tfrac{\Delta \bbra{\mu_5}}{k_B T} \bigr) \bigl( \tfrac{g_S}{106.75} \bigr)^{-1} 
    \;,
\end{split}
\end{equation}
\end{subequations}
where $g_S$ is the adiabatic number of degrees of freedom in the plasma at the time when the CPI develops.
For this estimate, we have assumed radiation domination with $g_E = g_S = 106.75$ effective relativistic degrees of freedom, we assume comoving entropy conservation between the electroweak epoch and today, and we denote the scale factor today as $a_0$.
We will consider these assumptions for all the remaining numerical estimates.
Note that to convert between energy density and G$^2$, we use the relation $\mathrm{G}^2 / 2 \approx 24.8 \GeV / \mathrm{cm}^3$ in \Eq{eq:delta_hM}.
For reference, the TeV blazar observations based on time-delayed $\gamma$-ray emission favor a nonzero magnetic field with  
$B^2 \lambda_B \gtrsim (1.8 \times10^{-17} \, \mathrm{G})^2 \, \mathrm{Mpc} \, a_0^3$; see the diagonal boundary on Fig.~5 of \rref{MAGIC:2022piy}.
For a maximally-helical magnetic field,
$|\Delta h_M| \approx B^2 \lambda_B$,
and otherwise the helicity is smaller.  
Therefore, as emphasized by the authors of \rref{Brandenburg:2017rcb}, the CPI is unable to generate a strong enough magnetic field to explain the blazar observations.  
The preceding discussion assumes that the chiral asymmetry is not being sourced.
In the next section, we discuss how the development of the CPI is affected by the presence of a source.  

\section{Development of CPI with a source}
\label{sec:CPI_w_source}

We are interested in a system with chiral sources, which cause the chiral chemical potential to grow in time.  
If the parameters are chosen such that $\Gamma_5^{-1} \ll t_\mathrm{cross} \ll t_\phi$, where $t_\mathrm{cross}$ is defined below, then the evolution with conformal time $t$ can be understood as follows.  
We assume that the comoving $\Gamma_5$ is constant in time during the sourcing process.  

\paragraph{$t_i < t < \Gamma_5^{-1} \mathrm{:}$}
Initially, the chemical potential rises from zero  $\propto t^2$, because
\begin{equation}
    \tilde{\mu}_5 (t) \approx
    \int_{t_i}^t \! \dd t' \tilde{S}_5 (t') \approx \sqrt{\erm} \, \tilde{\bar S}_5 
    t^2/(2 t_\phi)
    \;.
\end{equation}
This follows from \Eq{eq:dmudt} when $\tilde \mu_5 \approx 0$ and $\BB \approx 0$.

\paragraph{$\Gamma_5^{-1} < t < t_\mathrm{cross} \mathrm{:}$}
Subsequently the chemical potential rises $\propto t^1$ and tracks 
\begin{align}\label{eq:mu5_equilib}
    \tilde{\mu}_5(t) 
    \approx \tilde{S}_5(t) \, / \, \Gamma_5 
    \;,
\end{align}
where $\tilde{S}_5(t)$ is given by \Eq{eq:S5tilde}. 
Since chirality-violating reactions are in equilibrium on time scales that are longer than $\Gamma_5^{-1}$, the washout term in \Eq{eq:dmudt} would drive $\tilde{\mu}_5$ to zero if no sources were present.  
However, the nonzero source shifts the equilibrium point away from zero, and prevents the complete erasure of the chiral asymmetry.  

\paragraph{$t = t_\mathrm{cross} \mathrm{:}$}
While $\tilde{\mu}_5(t)$ is growing, the time scale for the CPI is decreasing
$t_\CPI{}(t) = 4 / [\eta k_\CPI{}^2(t)] = 4 / [\eta |\tilde{\mu}_5(t)|^2]$.
The CPI begins to develop when $t$ first exceeds $t_\CPI{}(t)$, which happens at a time 
\begin{align}\label{eq:tcross}
	t_\mathrm{cross} 
	= t_\CPI{}(t_\mathrm{cross}) 
	\approx \frac{2^{2/3} t_\phi^{2/3} \Gamma_5^{2/3}}{\tilde{\bar{S}}_5^{2/3} \eta^{1/3}} 
	\;.
\end{align}
The first equality defines $t_\mathrm{cross}$.  
The second equality was derived assuming $\Gamma_5^{-1} < t_\mathrm{cross} < t_\phi$, using \Eq{eq:mu5_equilib}, and approximating $\tilde S_5 \approx \sqrt{\erm} \, \tilde {\bar S}_5 \, t/t_\phi$.  
If instead $t_\mathrm{cross} > t_\phi$ then the source turns off before the CPI occurs, and there is no magnetic field amplification.  
Note that the instability scale is set by 
\begin{align}\label{eq:kcross}
    k_\mathrm{cross} 
    & = \tfrac{1}{2} k_\CPI{}(t_\mathrm{cross}) 
	\approx \frac{\tilde{\bar{S}}_5^{1/3}}{2^{1/3} t_\phi^{1/3} \Gamma_5^{1/3} \eta^{1/3}} 
	\;,
\end{align}
which represents the most rapidly growing mode at $t_\mathrm{cross}$.
The CPI leads to the amplification of a helical magnetic field from an initial seed field, as we discussed at \Eq{eq:Deltah}.  

\paragraph{$t_\mathrm{cross} < t < t_\phi \mathrm{:}$}
The source remains active while the CPI is developing.  
As a result we expect to see a sustained production of helical magnetic field across a range of scales.  
This is because the instability begins at time $t_\mathrm{cross}$ on scales set by $k_\mathrm{cross} = \tfrac{1}{2} \tilde{\mu}_5(t_\mathrm{cross})$, but soon afterward $\tilde{\mu}_5(t)$ grows $\propto t$ to track the increasing source \eqref{eq:mu5_equilib}, which shifts the CPI to larger $k = \tfrac{1}{2} \tilde{\mu}_5(t)$ and, hence, smaller length scales.  

\paragraph{$t = t_\phi \mathrm{:}$}
As the source reaches its maximum value $\tilde{S}_5(t_\phi) = \tilde{\bar{S}}_5$, the equilibrium chemical potential \eqref{eq:mu5_equilib} also reaches its maximum $\tilde{\mu}_5(t_\phi) \approx \tilde{\bar{S}}_5 / \Gamma_5$.  
At this time the CPI is active at the largest wavenumbers, with the fastest growth occurring at
\begin{align}\label{eq:kphi}
    k_\phi 
    = \tfrac{1}{2} k_\CPI{}(t_\phi) 
    = \tfrac{1}{2} \tilde{\mu}_5(t_\phi)
    \approx \tfrac{1}{2} \tilde{\bar{S}}_5 \, / \, \Gamma_5
    \;.
\end{align}
All in all, because the source grows the chiral chemical potential, the CPI instability scale scans across a range of wavenumbers $k_\mathrm{cross} < k < k_\phi$. 

\paragraph{$t_\phi < t \mathrm{:}$}
The source term becomes exponentially suppressed for $t \gtrsim t_\phi$.  
If no CPI has occurred, then $\tilde{\mu}_5(t)$ will track the equilibrium solution \eqref{eq:mu5_equilib} and drop to zero.  
However, if the CPI has taken place, then the system contains a helical magnetic field.  
The slowly-decaying helicity of the magnetic field also acts as a source of chirality~\cite{Fujita:2016igl,Kamada:2016eeb,Kamada:2016cnb,Hamada:2025cwu,Fukuda:2025nmc}, which corresponds to the Chern-Simons term in \Eq{eq:dmudt}, $\lambda c \EE \cdot \BB = \tfrac{1}{c} \lambda \eta \JJ \cdot \BB - \eta \tilde \mu_5 \BB^2$.
As a result, a new equilibrium solution is reached where
\begin{align}\label{eq:mu5_equilib_2}
    \tilde{\mu}_5(\xx,t) \approx &\, \frac{\lambda \eta}{\Gamma_5 + \lambda \eta \BB^2}
    \BB \cdot \nab \times \BB \\ \nonumber \approx &\,
    \frac{\lambda \eta}{\Gamma_5} \BB \cdot \nab \times \BB
    \;.
\end{align}
The second approximation is satisfied when $\lambda \eta |\BB|^2/\Gamma_5 \ll 1$.
As the magnetic field slowly decays, the chiral chemical potential relaxes toward zero.
In this phase, the chiral chemical potential becomes inhomogeneous due to the sourcing from the helical magnetic field.
In previous stages, since the source is assumed to be homogeneous, also $\tilde \mu_5$ remains approximately homogeneous.

In the remainder of this section we estimate the magnetic helicity that is expected to arise from these dynamics.  
Previously at \Eq{eq:Deltah} we presented a conservation law for magnetic helicity that was derived by setting $\tilde{S}_5 = \Gamma_5 = 0$.  
When either the source or washout are nonzero, the more general relation may be derived from the equation of motion for $\tilde \mu_5$ \eqref{eq:dmudt} by integrating over time and averaging over space.  
Doing so leads to the following conservation law,
\begin{equation}\label{eq:hel_prod}
    \Delta h_{\rm tot} (t) = \frac{2}{\lambda}
    \int_0^t \! \dd t^\prime \, \bigl[\tilde S_5(t') - \Gamma_5
    \langle \tilde \mu_5 \rangle(t^\prime)\bigr]\;,
\end{equation}
where $h_{\rm tot} = h_M + \tfrac{2}{\lambda} \langle \tilde \mu_5 \rangle$ is the total net helicity, composed of the magnetic helicity and the chiral chemical potential.

During the initial stage of the evolution, before $t_\mathrm{cross}$, the CPI has not yet kicked in and, hence, we do not expect significant evolution of the magnetic helicity but just fluctuations from the initial conditions.
However, during this period, we expect to have production of $h_\tot$ in the form of the chiral chemical potential, which will be eventually converted into magnetic helicity.
During the initial regime of the evolution $t_i < t < \Gamma_5^{-1}$,
\begin{equation}
    \Delta h_{\rm tot}(t < \Gamma_5^{-1}) 
    \approx \frac{2}{\lambda}
    \tilde \mu_5 (t) \approx \frac{\sqrt{\erm} \, \tilde{\bar S}_5 t^2}{\lambda t_\phi} 
    \;.
\end{equation}
At time $\Gamma_5^{-1} < t \lesssim t_\phi$, the chiral chemical potential is the result of a balance between the source and the washout terms, $\tilde \mu_5 (t) \approx \tilde S_5 (t)/\Gamma_5$ in \Eq{eq:mu5_equilib}.
During this period, the integrand in \Eq{eq:hel_prod} is close to zero.
However, it is not exactly zero, as we have a residual dependence from the equation of motion \eqref{eq:dmudt},
\begin{equation}
    \tilde S_5 - \Gamma_5 \langle \tilde \mu_5\rangle 
    \approx \tfrac{\partial}
    {\partial t} \langle \tilde \mu_5 \rangle + \lambda \eta
    \langle (\tilde \mu_5 \BB - \tfrac{1}{c}
    \JJ)\cdot \BB \rangle 
    \;.
\end{equation}
While the strength of the magnetic field is negligible,
\begin{equation}\label{creation_heli}
    \tilde S_5 - \Gamma_5 \langle \tilde \mu_5\rangle 
    \approx \tfrac{\partial}
    {\partial t} \langle \tilde \mu_5 \rangle 
    \;,
\end{equation}
which implies $\Delta h_{\rm tot} \approx \tfrac{2}{\lambda} \Delta \tilde \mu_5$ also during this stage.
Hence, as $\tilde \mu_5$ grows following the source, $h_\tot$ grows and, after the source (and $\tilde \mu_5$) reaches its maximum value, it starts to decrease.
Once the CPI starts at $t \gtrsim t_\mathrm{cross}$, the magnetic helicity grows exponentially and saturates shortly after $t_\phi$, at $t_{\rm sat} \gtrsim t_\phi$.
After this time, the chiral chemical potential balances with the  Chern-Simons term as described in \Eq{eq:mu5_equilib_2}.
Therefore, the saturated value of the magnetic helicity can be estimated as
\begin{align}
    h_M(t > t_{\rm sat}) \approx &\, \frac{\sqrt{\erm} \tilde {\bar S}_5}{\lambda t_\phi \Gamma_5^{2}} + \frac{2}{\lambda} \bigl[\tilde \mu_5 (t_{\rm sat}) - \tilde \mu_5 (\Gamma_5^{-1})
    \bigr] \nonumber \\ &\, - 2 \eta
    \int_{t_{\rm sat}}^t \dd t' \langle \BB \cdot \nab \times \BB \rangle
    \;,
\end{align}
where we have assumed that the condition $\lambda \eta |\BB|^2/\Gamma_5 \ll 1$ is satisfied.
If this is not the case, it is enough to include the additional term in the denominator in \Eq{eq:mu5_equilib_2}.
Furthermore, if the amplitude of the magnetic field resultant from the CPI is small, then the terms proportional to $\langle \BB \cdot \JJ \rangle$ can be neglected, leading to the following estimate of the magnetic helicity produced from the CPI with a source,
\begin{equation}\label{eq:hel_prod_source}
    h_M 
    \approx \frac{\sqrt{\erm} \tilde{\bar S}_5}{\lambda t_\phi \Gamma_5^2
    } \approx \frac{\sqrt{\erm}}{2 t_\phi \Gamma_5} \times \frac{2 \tilde{\bar S}_5}{\lambda \Gamma_5} 
    \;,
\end{equation}
where we have also assumed that $\tilde \mu_5 (\Gamma_5^{-1}) \approx \tilde\mu_5 (t_{\rm sat})$.
Note that the second equality expresses $h_M$ as a fraction $\sqrt{\erm}/(2 t_\phi \Gamma_5)$ of the magnetic helicity that would be produced in the absence of washout and source terms if the maximum chiral chemical potential at $t_\phi$ in our model would be converted into magnetic helicity.  
In the following sections we compare this analytical approximation with the result of direct numerical computation.

\section{Parameter estimates}
\label{sec:parameter}

\begin{table*}[t]
  \centering
  \caption{\label{tab:units}Units of temperature, time, length, and energy.
  }
  \vspace{0.2ex}
  \begin{tabular}{@{} l l l l @{}}
    \toprule
     \textbf{Name} & \textbf{Symbol} & \textbf{Relation} & \textbf{Value} \\
     \text{physical temperature} & $T_{\mathrm{phys},*}$ & ---& $100\,\mathrm{GeV}/k_B$ \\ 
     \text{comoving temperature} & $T_\ast$ & $a_\ast T_{\mathrm{phys},\ast}$ & $(7.77\times10^{-5}\,a_0\mathrm{eV}/k_B) \bigl( \tfrac{g_{E,\ast}}{106.75} \bigr)^{-1/3}$ \\
    \text{scale factor} & $a_\ast$ & $\bigl( \tfrac{g_{S,0}}{g_{S,\ast}} \bigr)^{1/3} \bigl( \tfrac{T_{\mathrm{phys},0}}{T_{\mathrm{phys},\ast}} \bigr)^{} a_0$ & $(7.77\times10^{-16}\,a_0) \bigl( \tfrac{g_{S,\ast}}{106.75} \bigr)^{-1/3} \bigl( \tfrac{T_{\mathrm{phys},*}}{100\,\mathrm{GeV}/k_B} \bigr)^{-1}$ \\
    \text{cosmic time} & $t_{c,*}$ & $1/(2 H_{\mathrm{phys},\ast})$ & $(2.34\times10^{-11}\,\mathrm{s}) \bigl( \tfrac{g_{E,\ast}}{106.75} \bigr)^{-1/2} \bigl( \tfrac{T_{\mathrm{phys},*}}{100\,\mathrm{GeV}/k_B} \bigr)^{-2}$ \\ 
    \text{conformal time} & $t_\ast$ & $2 t_{c,\ast} / a_\ast$ & $(6.03\times10^{4}\,\mathrm{s}/a_0) \bigl( \tfrac{g_{S,\ast}}{106.75} \bigr)^{1/3} \bigl( \tfrac{g_{E,\ast}}{106.75} \bigr)^{-1/2} \bigl( \tfrac{T_{\mathrm{phys},*}}{100\,\mathrm{GeV}/k_B} \bigr)^{-1}$ \\
     \text{physical length} & $l_{\mathrm{phys},\ast}$ & $c / H_{\mathrm{phys},\ast}$ & $(1.40\,\mathrm{cm}) \bigl( \tfrac{g_{E,\ast}}{106.75} \bigr)^{-1/2} \bigl( \tfrac{T_{\mathrm{phys},*}}{100\,\mathrm{GeV}/k_B} \bigr)^{-2}$ \\ 
    \text{comoving length} & $l_\ast$ & $l_{\mathrm{phys},\ast} / a_\ast$ & $(1.81\times10^{15}\,\mathrm{cm}/a_0) \bigl( \tfrac{g_{S,\ast}}{106.75} \bigr)^{1/3} \bigl( \tfrac{g_{E,\ast}}{106.75} \bigr)^{-1/2} \bigl( \tfrac{T_{\mathrm{phys},*}}{100\,\mathrm{GeV}/k_B} \bigr)^{-1}$ \\
    \text{physical energy} & $E_{\mathrm{phys},\ast}$ & $\rho_{\mathrm{phys},\ast} l_{\mathrm{phys},\ast}^3$ & $(1.27\times10^{60}\,\mathrm{eV}) \bigl( \tfrac{g_{E,\ast}}{106.75} \bigr)^{-1/2} \bigl( \tfrac{T_{\mathrm{phys},*}}{100\,\mathrm{GeV}/k_B} \bigr)^{-2}$ \\
    \text{comoving energy} & $E_\ast$ & $a_\ast E_{\mathrm{phys},\ast}$ & $(9.84\times10^{44}\,a_0\mathrm{eV}) \bigl( \tfrac{g_{S,\ast}}{106.75} \bigr)^{-1/3} \bigl( \tfrac{g_{E,\ast}}{106.75} \bigr)^{-1/2} \bigl( \tfrac{T_{\mathrm{phys},*}}{100\,\mathrm{GeV}/k_B} \bigr)^{-3}$ 
  \end{tabular}
\end{table*}

In this section, we discuss the parameters of sourced \cMHD{} \eqref{eq:cMHD}, 
assuming that the source $S_5(t)$ has the functional form shown in \Eq{eq:S5}.  
We also have the relations $\eta = c^2 / \sigma$, $\alpha = e^2 / (4 \pi \hbar c) \approx 1/137$, the expression for $\tilde{\bar{S}}_5$ from \Eq{eq:S5tilde}, and the expression for $\lambda_\star$ from \Eq{eq:lambda}.
To provide numerical estimates of the parameters, we suppose that the source is active during the electroweak epoch in the early universe, just before the electroweak phase transition.  
Then dimensionful quantities are expressed using the units shown in \tref{tab:units}.  
In a radiation-dominated universe with physical temperature $T_{\mathrm{phys},\ast}$, $g_{E,\ast}$ effective relativistic degrees of freedom, and $g_{S,\ast}$ effective adiabatic degrees of freedom, the average energy density is $\rho_{\mathrm{phys},\ast} = k_B^4 \pi^2 g_{E,\ast} T_{\mathrm{phys},\ast}^4 / (30 \hbar^3 c^3)$ and the Hubble expansion rate is $H_{\mathrm{phys},\ast} = [\hbar \rho_{\mathrm{phys},\ast} / (3 c \Mpl^2)]^{1/2}$.  
Our units of time, length, and energy are chosen to be a Hubble time, a Hubble length, and a Hubble energy.

It is useful to note the conversion $\tilde{\mu}_5 = 2 \alpha \mu_5 / (\pi \hbar c)$.  
In terms of the fiducial length and temperature, this implies
\begin{align}\label{eq:mu5_convert}
    \tfrac{\mu_5}{k_B T_\ast} 
    \approx &\,
    \bigl( 3.02 \times 10^{-14} \bigr) 
    \bigl( \tfrac{\tilde{\mu}_5}{l_\ast^{-1}} \bigr) \\ \nonumber & \times 
    \bigl( \tfrac{T_{\phys, \ast}}{100 \GeV/k_B} \bigr)
    \bigl( \tfrac{g_E}{106.75} \bigr)^{1/2}
    \;.
\end{align}
Our calculation assumes $\mu_5 / (k_B T_\ast) \ll 1$, which corresponds to $\tilde{\mu}_5 / l_\ast^{-1} \ll 10^{14}$ for the fiducial parameters.
Prior to the electroweak phase transition, it is more accurate to talk about the hypermagnetic field rather than the electromagnetic field, and the chiral asymmetry is only carried by the right-chiral electrons and left-chiral positrons.  
However, these subtleties only impact our calculations modulo order one factors, and we neglect them going forward.  
See \rref{Brandenburg:2023imm} for additional discussion.

One advantage of working in a plasma at (or above) the electroweak scale is that the electroweak symmetry is unbroken, and since all particles are massless, the various properties of the plasma are only controlled by the plasma temperature.  
We take $D_5 = \nu = \eta$ for simplicity, and we adopt the approximation ${\sigma_\star} \approx \tfrac{k_B}{\hbar} (1 \times 10^2) T_\ast$ \cite{Arnold:2000dr}.
We calculate $\lambda_\star \approx \tfrac{\hbar c}{k_B^2} (6.5 \times 10^{-5}) T_\ast^{-2}$ using \Eq{eq:lambda}.  
Scattering reactions that lead to chiral charge erasure are mediated by the electron's Yukawa interaction.  
Based on \rref{Bodeker:2019ajh}, we estimate the chirality erasure rate as $\Gamma_{5, \star} \approx \tfrac{k_B}{\hbar} (1.3 \times 10^{-2}) y_e^2 T_\ast$ where $y_e \approx 2.94 \times 10^{-6}$ is the electron Yukawa coupling.  
These estimates are summarized in the ``Fiducial'' row of \tref{tab:runs}, where we omit the dependency on $T_\ast$, $g_E$, and $g_S$ for simplicity.
Note that the product $\lambda \eta |\BB|^2 / \Gamma_5$ appearing in \Eq{eq:mu5_equilib_2} is 
\begin{align}
    \tfrac{\lambda_\star \eta_\star B^2}{\Gamma_{5, \star}} 
    & \approx \tfrac{\hbar^3 c^3}{k_B^4} (6.5 \times 10^{-5}) y_e^{-2}  T_\ast^{-4} B^2
    \\ 
    & \approx \bigl( 4.1 \times 10^8 \, \Omega_B \bigr) \bigl(\tfrac{g_E}{106.75}\bigr) 
    \;,
    \nonumber 
\end{align}
where $\Omega_B = \half B^2/(E_\ast l_\ast^{-3})$.

The expected magnetic helicity is given by \Eq{eq:hel_prod_source}, and evaluating this formula for the fiducial parameters: 
\begin{align}
    \Delta h_M \approx &\, (1.5 \times 10^{-19} \G)^2 \Mpc \, a_0^3 
    \bigl( \tfrac{t_{\phi, \star}}{t_\phi} \bigr)
    \bigl( \tfrac{\Gamma_{5, \star}}{\Gamma_5} \bigr) 
    \nonumber \\ &\, 
    \times
    \bigl( \tfrac{\lambda_\star}{\lambda} \bigr) 
    \bigl( \tfrac{{\mu_5}(t_\phi)}{k_B T_\ast} \bigr) 
    \bigl( \tfrac{g_S}{106.75} \bigr)^{-1} 
    \;.
\end{align}
This is slightly smaller than the expected magnetic helicity calculated using \Eq{eq:Deltah}, which ignores the source and washout terms.  
However, we note that the inclusion of a sourcing mechanism allows to justify that the CPI arises in the electroweak epoch around $100 \GeV$.

The amplitude of the chiral source is given by \Eq{eq:S5tilde}, and evaluating this formula for the fiducial parameters: 
\begin{align}
    \tilde{\bar S}_{5, \star} \approx &\, (1.62 \times 10^{13} l_\ast^{-1}
    t_\ast^{-1}) \bigl(\tfrac{g_E}{106.75}\bigr)^{1/2} \bigl(\tfrac{\epsilon \beta \Omega_\phi}{10^{-5}} \bigr) \\ \nonumber &\, \times \bigl(\tfrac{m_\phi c^2}{100 \GeV}
    \bigr)^{-1} \bigl(\tfrac{t_\phi}{t_{\phi, \star}}\bigr)^{-2} 
    \;.
\end{align}
Note that the free parameters are $t_\phi$, $m_\phi$, $\epsilon$, $\beta$, and $\Omega_\phi$, but they only contribute to the source through the product $\tilde{\bar{S}}_5 \propto t_\phi^2 \epsilon \beta \Omega_\phi / m_\phi$. 
To obtain a reasonable estimate for the source, we fiducialize to $\epsilon \beta \Omega_\phi = 10^{-5}$, $m_\phi \approx k_B T_{\phys, \ast}/c^2 = 100 \GeV/c^2$, and $t_{\phi, \star} = 0.05 \, t_\ast$. 
An arbitrarily smaller source is also possible (e.g., if $|\epsilon| \ll 1$).  
On the other hand, a much larger source would come into conflict with our assumption that the chemical potential is small.  
The condition $\mu_5/(k_B T_\ast) \ll 1$ implies $\tilde \mu_5/l_\ast^{-1} \ll 3 \times 10^{-14}$, $\tilde{\bar S}_5 \ll 2.6 \times 10^{16} l_\ast^{-1} t_\ast^{-1}$, and $\epsilon \beta \Omega_\phi \ll 10^{-2}$ via Eqs.~\eqref{eq:S5tilde},~\eqref{eq:mu5_equilib},~and~\eqref{eq:mu5_convert}. 

The fiducial time scales are estimated to be 
\begin{subequations}\label{eq:time_scales}
\begin{align}
    \Gamma_5^{-1} & = 
    (1.25 \times 10^{-3} \, t_\ast)
    \bigl( \tfrac{\Gamma_5}{\Gamma_{5,\star}} \bigr)^{\!-1} 
    \;, \\ 
    t_\mathrm{cross} & \approx
    (2.6 \times 10^{-2} \, t_\ast)
    \bigl( \tfrac{t_\phi}{t_{\phi, \star}} \bigr)^{2/3} 
    \bigl( \tfrac{\Gamma_5}{\Gamma_{5,\star}} \bigr)^{2/3} 
    \\ & \qquad \times 
    \bigl( \tfrac{\eta}{\eta_\star} \bigr)^{\!-1/3} 
    \bigl( \tfrac{\tilde{\bar{S}}_5}{\tilde{\bar{S}}_{5,\star}} \bigr)^{-2/3} 
    \;, \nonumber \\ 
    t_\phi & = (0.05 \, t_\ast)
    \bigl( \tfrac{t_\phi}{t_{\phi,\star}} \bigr)^{} 
    \;,
\end{align}
\end{subequations}
where we have used the formula for $t_\mathrm{cross}$ in \Eq{eq:tcross}.  
For our numerical simulations, we would like to have the time scales well separated and ordered such that $\Gamma_5^{-1} \ll t_\mathrm{cross} \ll t_\phi$.  
For simulation runs {\bf A} and {\bf A}$^\ast$ in \tref{tab:runs}, which start at time $t_i = 10^{-5} t_\ast$ and end at $t_f = 2 t_\ast$, the time scales are $t_i \ll \Gamma_5^{-1} = 10^{-3} t_\ast \ll t_\phi = 0.05 t_\ast \ll t_f$.

\begin{table}[t]
\centering
\resizebox{0.45\textwidth}{!}{
\begin{tabular}{|l|c|c|c|c|c|}
\hline
\textbf{Label} & $\eta$ = $D_5$ = $\nu$ & $\lambda$ & $\Gamma_5$ & $\tilde{\bar{S}}_5$ & $t_\phi$ \\
\hline
Units & $l_\ast^2 t_\ast^{-1}$ & $l_\ast E_\ast^{-1}$ & $t_\ast^{-1}$ & $l_\ast^{-1} t_\ast^{-1}$ & $t_\ast$ \\ 
\hline
Fiducial &
$1.4\times 10^{-18}$ &
$1.2\times 10^{29}$ &
$8.0\times 10^{2}$ &
$1.6 \times 10^{13}$ &
$0.05$ \\
\hline\hline
\textbf{A} & $10^{-6}$ & $10^{8}$ & $10^{3}$ & $4\times10^{7}$ & $0.05$ \\ 
\hline
$\textbf{A}^*$ & $10^{-6}$ & $10^{8}$ & $10^{3}$ & $4\times10^{7}$ & $0.05$ \\ 
\hline
\textbf{B} & $10^{-6}$ & $1.6\times10^{9}$ & $10^{3}$ & $4\times10^{7}$ & $0.05$ \\
\hline
\textbf{C} & $10^{-6}$ & $1.6\times10^{10}$ & $10^{3}$ & $4\times10^{7}$ & $0.05$ \\
\hline
\textbf{D} & $10^{-6}$ & $10^{12}$ & $10^{3}$ & $4\times10^{7}$ & $0.05$ \\ 
\hline
\end{tabular}
}
\caption{\label{tab:runs}
Parameter values. Units are multiples of the comoving Hubble energy and length, $E_\ast$ and $l_\ast$, and conformal time, $t_\ast$.  Fiducial values are present to make a conversion from code units to physical ones.  Simulations use $N = 256$ sites ($1024$ for $\textbf{A}^\ast$) with periodic boundary conditions, corresponding to a range of wavenumbers from $k_\mathrm{min} = 500 \, l_\ast^{-1}$ ($200 \, l_\ast^{-1}$ for $\textbf{A}^\ast$) to $k_\mathrm{max} = 5.0 \times 10^3 \, l_\ast^{-1}$ ($1.0 \times 10^5 \, l_\ast^{-1}$ for $\textbf{A}^\ast$). 
}
\end{table}

The fiducial length scales are estimated as follows.  
We suppose that the equations of \cMHD{} are solved on a three-dimensional cubic lattice with $N$ sites per side spaced uniformly in intervals of $\delta x$,
spanning a distance $N \delta x$. 
We identify the corresponding wavenumbers as $k_\mathrm{min} = 2 \pi / (N \delta x)$ and $k_\mathrm{max} = N k_\mathrm{min} / 2 = \pi / \delta x$.   
Using the fiducial parameter values in \tref{tab:runs} and the formulas for $k_\mathrm{cross}$ and $k_\phi$ in \Eqs{eq:kcross}{eq:kphi}, we estimate the wavenumbers:
\begin{subequations}\label{eq:wavenumbers}
\begin{align}
    k_\mathrm{min} & = (2.0 \times 10^2 \, l_\ast^{-1}) \bigl( \tfrac{k_\mathrm{min}}{200 \, l_\ast^{-1}} \bigr)\;, \\ 
    k_\mathrm{cross} & \approx
    (5.2 \times 10^9 \, l_\ast^{-1}) \bigl( \tfrac{t_\phi}{t_{\phi,\star}} \bigr)^{\!-1/3} \bigl( \tfrac{\Gamma_5}{\Gamma_{5,\star}} \bigr)^{\!-1/3} \\ 
    & \qquad \times \bigl( \tfrac{\eta}{\eta_\star} \bigr)^{\!-1/3} \bigl( \tfrac{\tilde{\bar{S}}_5}{\tilde{\bar{S}}_{5,\star}} \bigr)^{\!1/3} \;,\nonumber \\ 
    k_\phi & \approx
    (1.0 \times 10^{10} \, l_\ast^{-1}) \bigl( \tfrac{\Gamma_5}{\Gamma_{5,\star}} \bigr)^{\!-1} \bigl( \tfrac{\tilde{\bar{S}}_5}{\tilde{\bar{S}}_{5,\star}} \bigr)^{}\;, \\ 
    k_\mathrm{max} & \approx (1.0 \times 10^5 \, l_\ast^{-1}) \, \bigl( \tfrac{N}{1024} \bigr) \bigl( \tfrac{k_\mathrm{min}}{200 \, l_\ast^{-1}} \bigr) 
    \;.
\end{align}
\end{subequations}
For our numerical simulations, we would like to have the length scales well separated and ordered such that $k_\mathrm{min} \ll k_\mathrm{cross} \ll k_\phi \ll k_\mathrm{max}$.
For simulation runs {\bf A} and {\bf A$^\ast$} in \tref{tab:runs}, we estimate $k_\mathrm{cross} \approx 7.4 \times 10^3 \, l_\ast^{-1}$ and $k_\phi \approx 2 \times 10^4 \, l_\ast^{-1}$.
Both length scales are well resolved for the simulations with $N = 256$ and $N = 1024$, for which $k_\mathrm{max} = 5 \times 10^3$ and $1.0  \times 10^5 \, l_\ast^{-1}$, respectively.

\section{Numerical \cMHD{} simulations}
\label{sec:numerical}

We solve \Eqss{eq:dAdt}{eq:dlnrhodt} using the {\sc Pencil Code} \cite{PencilCode:2020eyn}, which is a massively parallel code using sixth-order finite differences and a third-order time stepping scheme that can be used for MHD simulations in an expanding background.
Fields are discretized on a three-dimensional spatial lattice and evolved with time in steps of $\delta t$, which is determined adaptively.  
For most figures presented in this section, we use the parameter set $\textbf{A}^\ast$ in \tref{tab:runs}, corresponding to simulations with 1024$^3$ spatial grid points.

In order to solve \Eq{eq:cMHD} we specify initial conditions at the initial time $t_i = 10^{-5} \, t_\ast$.
These are chosen as 
\begin{subequations}\label{eq:initial_conditions}
\begin{align}
    \AAA(\xx,t_i) & = \mathcal{N}(10^{-12} \, E_\ast^{1/2} l_\ast^{-1/2})\;, \\ 
    \tilde{\mu}_5(\xx,t_i) & = 0\;,\\ 
     \uu(\xx,t_i) & = 0\;, \\ 
    \rho(\xx,t_i) & = E_\ast l_\ast^{-3} 
    \;.
\end{align}
\end{subequations}
At each point in space, the Cartesian components of the magnetic vector potential $\AAA = (A_x, A_y, A_z)$ are each drawn independently from a normal distribution with mean equal to zero and standard deviation shown in \Eq{eq:initial_conditions}.
This white noise corresponds to a power spectrum $\dd \rho_B / \dd \ln k = \tfrac{k^3}{2\pi^2} \tfrac{1}{2} P_{B}(k) \propto k^5$ for the magnetic field and $\tfrac{k^3}{2\pi^2} P_{A}(k) = k^2 \tfrac{k^3}{2\pi^2} P_{B}(k) \propto k^3$ for the magnetic potential.
The chiral chemical potential $\tilde{\mu}_5$ is taken to be homogeneous and equal to zero at the initial time, because the washout term would send it to zero before the source begins growing around the start of the simulation. 
The fluid velocity $\uu$ is taken to be homogeneous and equal to zero at the initial time, because the plasma would not be turbulent before the onset of the CPI.  
The fluid energy density $\rho$ is taken to be homogeneous and equal to the cosmological energy density initially. 

\begin{figure}[t]
    \centering
    \includegraphics[width=0.95\linewidth]{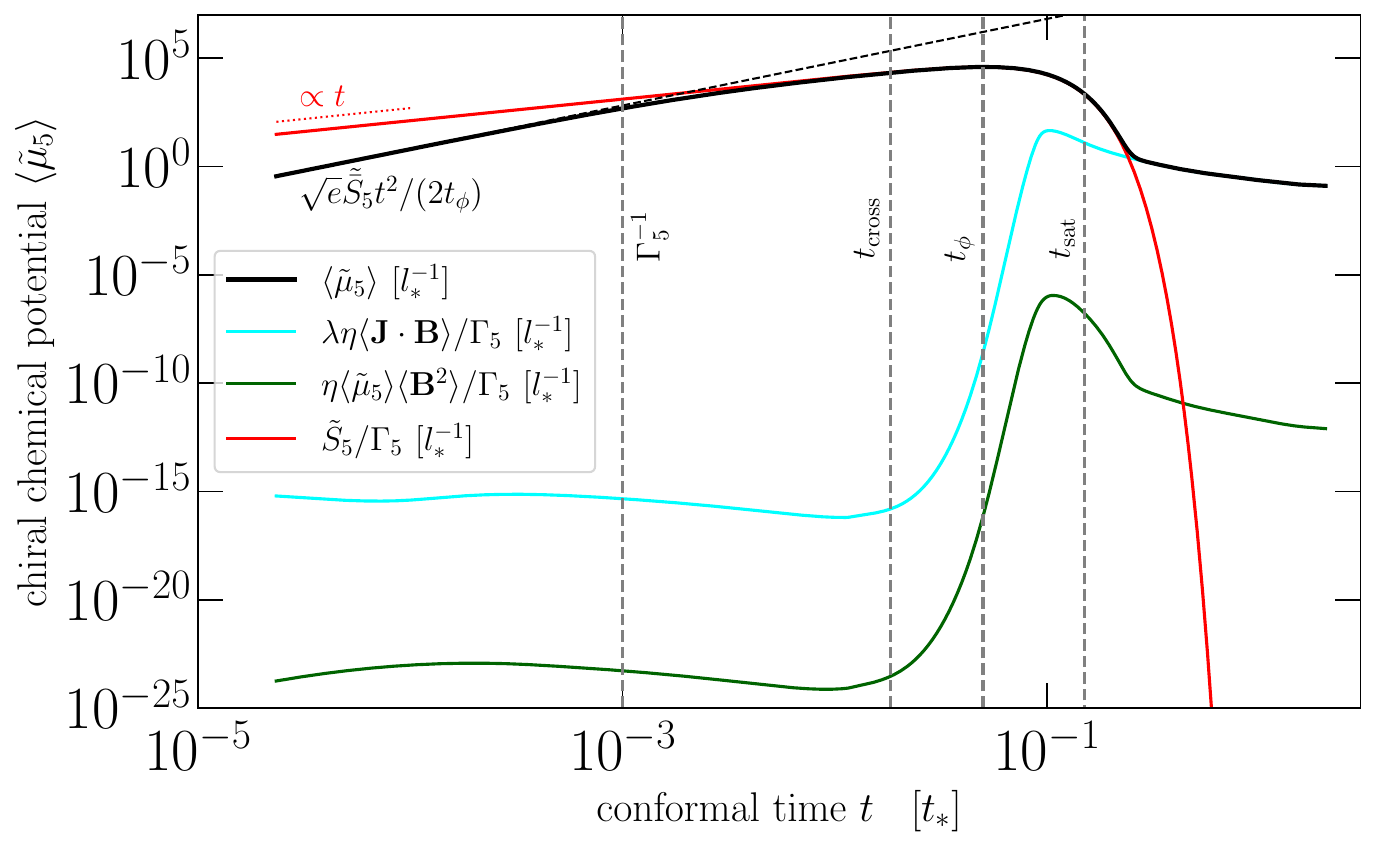}
    \caption{\label{fig:evolution}
    Evolution of the volume averaged chiral asymmetry.  We plot the comoving scaled volume averaged chiral chemical potential $\langle \tilde{\mu}_5(\xx,t) \rangle$ in units of $l_\ast^{-1}$ as a function of conformal time $t$ in units of $t_\ast$.  The chemical potential is calculated numerically by solving the equations of \cMHD{} for the run $\textbf{A}^\ast$ (see \tref{tab:runs}).  For comparison, we also plot the comoving chiral source $\tilde{S}_5(t) / \Gamma_5$, and the Chern-Simons terms $\eta \langle \tilde \mu_5 \rangle \langle \mathbf{B}^2 \rangle / \Gamma_5$ and $\eta \lambda \langle \JJ \cdot \BB \rangle / \Gamma_5$, in units of $l_\ast^{-1}$. Vertical dotted lines indicate the time scales in \Eq{eq:time_scales}. 
    }
\end{figure}

\Fref{fig:evolution} shows the time evolution of the chiral chemical potential $\langle \tilde{\mu}_5(t) \rangle$.  
For reference, we also show the chiral source $\tilde{S}_5(t)$, which is given by the formula in \Eq{eq:S5tilde}, divided by $\Gamma_5$.
At early times $t$, such that $t_i \leq t \lesssim \Gamma_5^{-1}$, the chemical potential rises from zero like $\propto t^2$, because $\tfrac{\partial}{\partial t} \tilde{\mu}_5 \sim \tilde{S}_5 \propto t$.  
At later times $t$, such that $\Gamma_5^{-1} \lesssim t \lesssim t_\mathrm{cross}$, the chemical potential rises like $\propto t^1$ and tracks $\tilde{\mu}_5 \approx \tilde{S}_5 / \Gamma_5$.  
The CPI begins at $t \approx t_\mathrm{cross}$, but the chemical potential continues to grow, because the source remains active until time $t_\phi$.  
After this time, the source begins to drop to zero exponentially quickly, and the chemical potential begins to track $\JJ \cdot \BB \propto \BB \cdot \nab \times \BB$, according to \Eq{eq:mu5_equilib_2}.
Note that for this simulation, the contribution to the asymptotic value of $\tilde \mu_5$ from the Chern-Simons term proportional to $\tilde \mu_5 |\BB|^2$ is $\lambda \eta/\Gamma_5 = 0.1 \, |\BB^2| / (E_\ast l_\ast^{-3}) \ll 1$.
For larger values of this ratio, $\tilde \mu_5$ would follow the combination given in \Eq{eq:mu5_equilib_2}.
At late times, the slowly decaying magnetic helicity continues to source the chiral asymmetry \cite{Boyarsky:2011uy}. 
The chiral chemical potential $\mu_5$ with units of energy may be calculated using the conversion in \Eq{eq:mu5_convert}.  

\begin{figure}[t]
    \centering
    \hfill
    \includegraphics[width=0.95\linewidth]{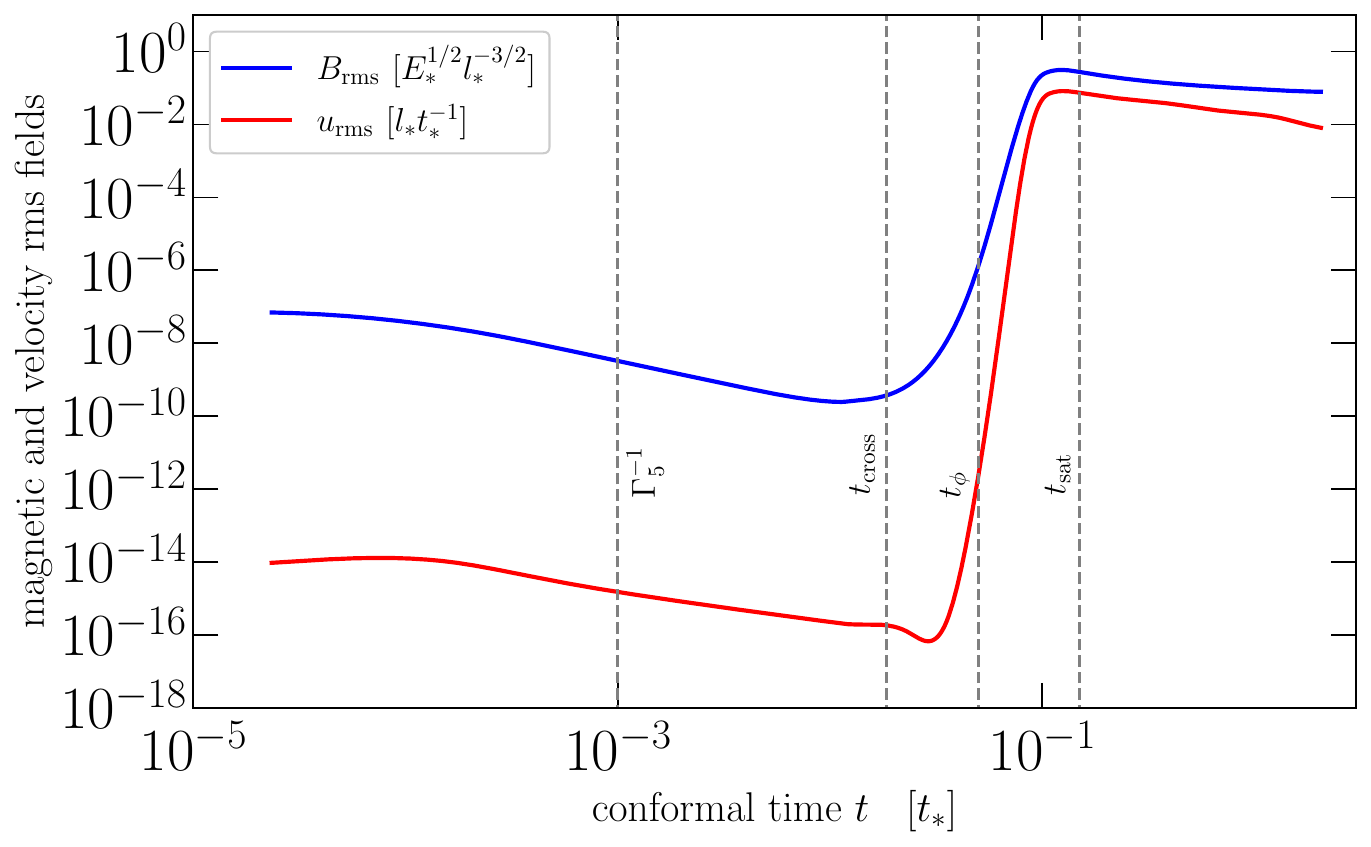}
    \caption{\label{fig:magnetic}
    Evolution of the root mean squared (rms) magnetic and velocity fields.  We plot the comoving rms field strength $B_\mathrm{rms}(t)$ in units of $E_\ast^{1/2} l_\ast^{-3/2}$
    and $u_{\mathrm{rms}}(t)$ in units of $l_{\ast} t_{\ast}^{-1}$
    as functions of conformal time $t$ in units of $t_\ast$.  Vertical dotted lines represent the time scales in \Eq{eq:time_scales}.  
    }
\end{figure}

\Fref{fig:magnetic} shows the evolution of the root mean square (rms) magnetic and velocity field strengths $B_\mathrm{rms}(t) = \langle \BB^2 (\xx, t) \rangle^{1/2}$
and $u_{\rm rms} (t) = \langle \uu^2 (\xx, t) \rangle^{1/2}$.
The initial white noise \eqref{eq:initial_conditions} sets the starting value of $B_\mathrm{rms} \sim k_\mathrm{max} A_\mathrm{rms} \sim 10^{-6} \, E_\ast^{1/2} l_\ast^{-3/2}$.  
Initially, when the chiral chemical potential and fluid velocity are negligibly small, the magnetic field strength decreases as a consequence of the magnetic diffusivity.  
This can be understood from \Eq{eq:A_evolution}, which leads to $A_{\kk,\pm} \propto \mathrm{e}^{- \eta |\kk|^2 t}$, and by noting that $1/(\eta k_\mathrm{max}^2) \approx 10^{-6} \, t_\ast^{}$ for the parameters in \tref{tab:runs}.  
The high-$k$ part of the white-noise spectrum is dissipated most quickly, leaving the lower-$k$ spectrum intact.  
When the time reaches $t \approx t_\mathrm{cross} \approx 0.018 \, t_\ast$, the CPI begins to develop and the magnetic field strength grows very quickly.  
This growth continues through the time $t = t_\phi = 0.05 \, t_\ast$ when the source reaches its largest value.  
Afterward the source begins to drop exponentially, the CPI ceases, and the magnetic field strength decreases.
The velocity field is driven by the Lorentz force until it reaches equipartition with the magnetic field at the end of the CPI.  
At later times, the velocity field decays as a power law in time following MHD turbulent decay.   

\begin{figure}[t]
    \centering
    \includegraphics[width=0.95\linewidth]{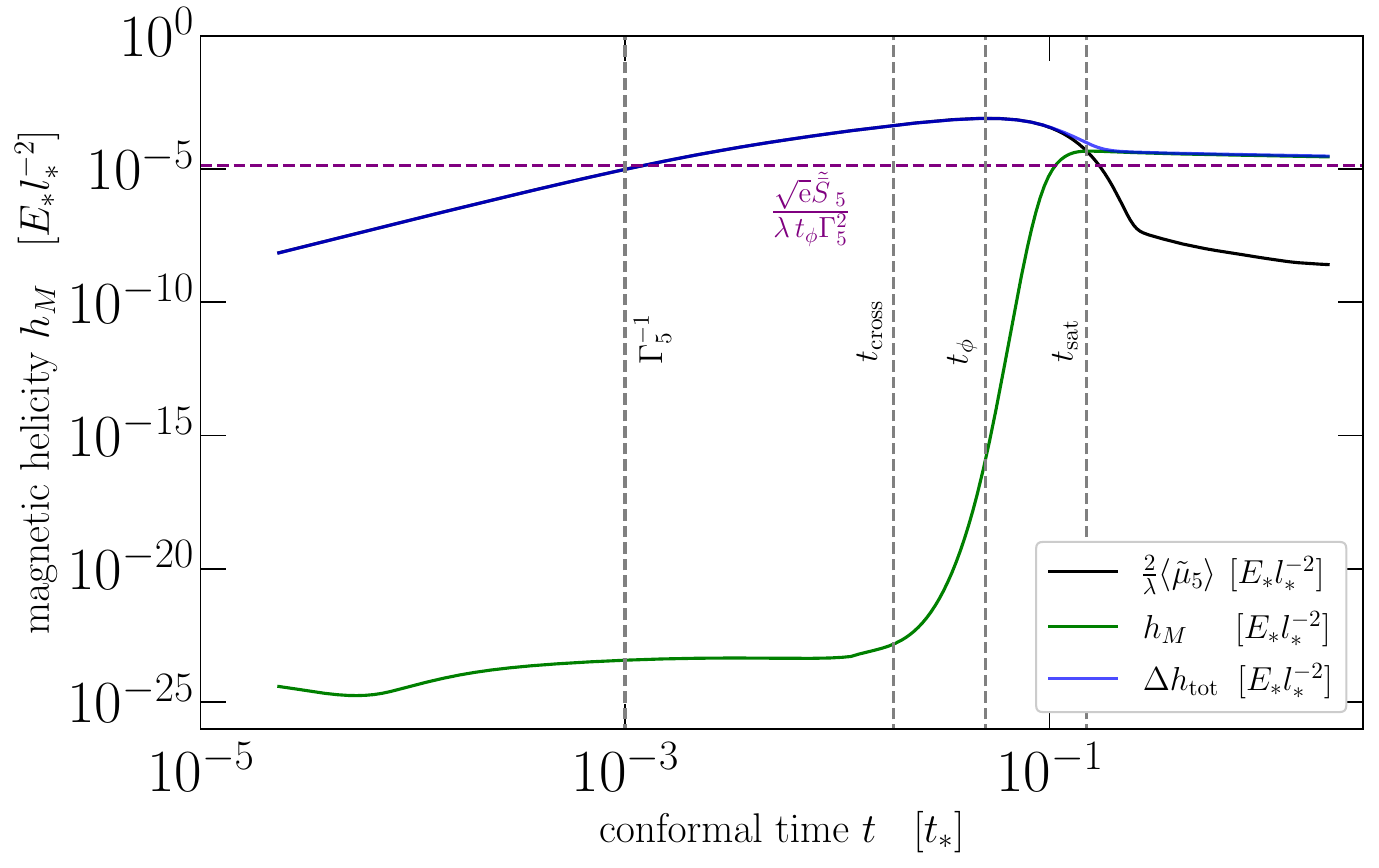}
    \hfill
    \caption{\label{fig:helicity}
    Evolution of the magnetic helicity.  We show the magnetic helicity $h_M$, the chiral chemical potential expressed in units of helicity, and the integral leading to a change in the total net helicity, $\Delta h_\tot$, where $h_\tot = h_M + \tfrac{2}{\lambda} \langle \tilde \mu_5 \rangle$.  The horizontal dotted line corresponds to the estimate of the magnetic helicity produced by the CPI with a source given in \Eq{eq:hel_prod_source}.  The numerical relative error $|h_\tot - \Delta h_\tot|/|h_\tot| < 10 \%$ at all times in the simulation, where $\Delta h_\tot$ corresponds to the integral over conformal time given in \Eq{eq:hel_prod}. Vertical dotted lines represent the time scales in \Eq{eq:time_scales}.  
    }
\end{figure}

\Fref{fig:helicity} shows the evolution of the total net helicity $h_\tot$ and each of its contributions: the magnetic helicity $h_M$ and the helicity associated to the chiral chemical potential, $\tfrac{2}{\lambda} \langle \tilde \mu_5 \rangle$.
Before the CPI $h_M$ is small, because the initial condition has zero average helicity, and $h_\tot$ is dominated by $\tilde \mu_5$.
At times $t = \Gamma^{-1}_5$, even though $\tilde{\mu}_5 \approx \tilde{S}_5/\Gamma_5$, $h_\tot$ keeps growing with time up to $t_\phi$, as discussed around \Eq{creation_heli}.
After this time, $h_\tot$ decreases until it equilibrates with the produced magnetic helicity at around $t_\sat$.
We find numerically that $\Delta h_\tot$, given by \Eq{eq:hel_prod}, is conserved in the simulations with a relative error smaller than 10\%. 

When the CPI saturates, the magnetic field and fluid begin to evolve together into a turbulent inverse cascade. 
During the inverse cascade, the comoving magnetic helicity density is approximately constant.
This time is indicated in the figures as $t_{\rm sat} \approx 0.15 t_\ast$.
This evolution is expected to continue until recombination when the plasma converts into neutral hydrogen, and afterward the comoving helicity density of the freely-expanding magnetic field is also constant.  
Consequently, the comoving magnetic helicity density today is approximately equal to the value it reaches after the CPI saturates.  
To convert between them, we use the relation $E_\ast / l_\ast^2 \approx 1.97 \times 10^{-21} \, a_0^3 \mathrm{G}^2 \mathrm{Mpc}$, where $\mathrm{G}^2 / 2 \approx 24.8 \GeV / \mathrm{cm}^3$.
The numerical value of the (roughly constant) helicity at the end of the simulation (see \fref{fig:magnetic}) is $h_M = 2.89 \times 10 ^{-5} \, E_\ast l_\ast^{-2}$, which implies: 
\begin{align}\label{eq:hM_numerical}
    h_M & \approx
    (2.4 \times 10^{-13} \G)^2 \Mpc \, a_0^3 
    \;.
\end{align}
This agrees well with our analytical estimates.
At \Eq{eq:Deltah} we gave the magnetic helicity that results from the CPI with a preexisting chiral asymmetry, assuming that the source and washout can be neglected.  
By setting $\tilde{\mu}_5 \approx 4 \times 10^4 \, l_\ast^{-1}$, which is the maximum value seen on \fref{fig:evolution}, and using the conversion in \Eq{eq:mu5_convert}, we estimate $h_M \approx (1.3 \times 10^{-12} \G)^2 \Mpc \, a_0^3$.
This is a factor of $\sim 30$ larger than the numerical result \eqref{eq:hM_numerical}, which is partly due to neglecting the washout.  
At \Eq{eq:hel_prod_source} we derived an approximation to the magnetic helicity that results from the CPI in the presence of source and washout.  
Evaluating that expression gives $h_M \approx (1.6 \times 10^{-13} \G)^2 \Mpc \, a_0^3$, which is in excellent agreement with the numerical result \eqref{eq:hM_numerical}.  
This agreement gives us confidence in the approximations used to derive \Eq{eq:hel_prod_source}.

Since the next two figures show spectra, let us first establish our conventions.  
The magnetic energy spectrum $\dd\rho_B / \dd\ln k$ is defined such that 
\begin{align}
    \rho_B (t) 
    & = \half \langle \BB^2 (\xx, t)\rangle 
    = \int_0^\infty \! \frac{\dd k}{k} \, \frac{\dd \rho_B}{\dd \ln k}
    \;,
\end{align}
and the magnetic helicity spectrum $\dd h_M/\dd \ln k$ is defined such that 
\begin{align}
    h_M (t) 
    & = \langle \AAA(\xx,t) \cdot \BB(\xx,t) \rangle 
    = \int_0^\infty \! \frac{\dd k}{k} \, \frac{\dd h_M}{\dd \ln k}
    \;, 
\end{align}
where $k = |\kk|$.  
These spectra are related to the power spectra used in \rref{Durrer:2013pga} by the relations $P_B(k, t) =  (2 \pi^2/k^3) \, \dd \rho_B/\dd \ln k$ and $P_{aB} (k, t) = (\pi^2/k^2) \, \dd h_M/\dd \ln k$.  
The helicity spectrum is bounded by the energy spectrum via the realizability condition $-P_B \leq P_{aB} \leq P_B$, which implies $- \tfrac{\dd\rho_B}{\dd \ln k} \leq \tfrac{k}{2} \tfrac{\dd h_M}{\dd \ln k} \leq \tfrac{\dd\rho_B}{\dd \ln k}$, and we define the helicity fraction $\varepsilon_H = \bigl( \tfrac{k}{2} \tfrac{\dd h_M}{\dd \ln k} \bigr) \bigl( \tfrac{\dd\rho_B}{\dd \ln k} \bigr)^{-1}$, which is bounded by $-1 \leq \varepsilon_H \leq 1$.
To compare with the conventions in other work, e.g. \cite{Brandenburg:2017rcb,Brandenburg:2017neh,Brandenburg:2023imm}, note that $E_M(k,t) = k^{-1} \dd\rho_B/\dd\ln k$ and $H_M(k,t) = k^{-1} \dd h_M/\dd\ln k$.  

\begin{figure}[t]
    \centering
    \includegraphics[width=0.95\linewidth]{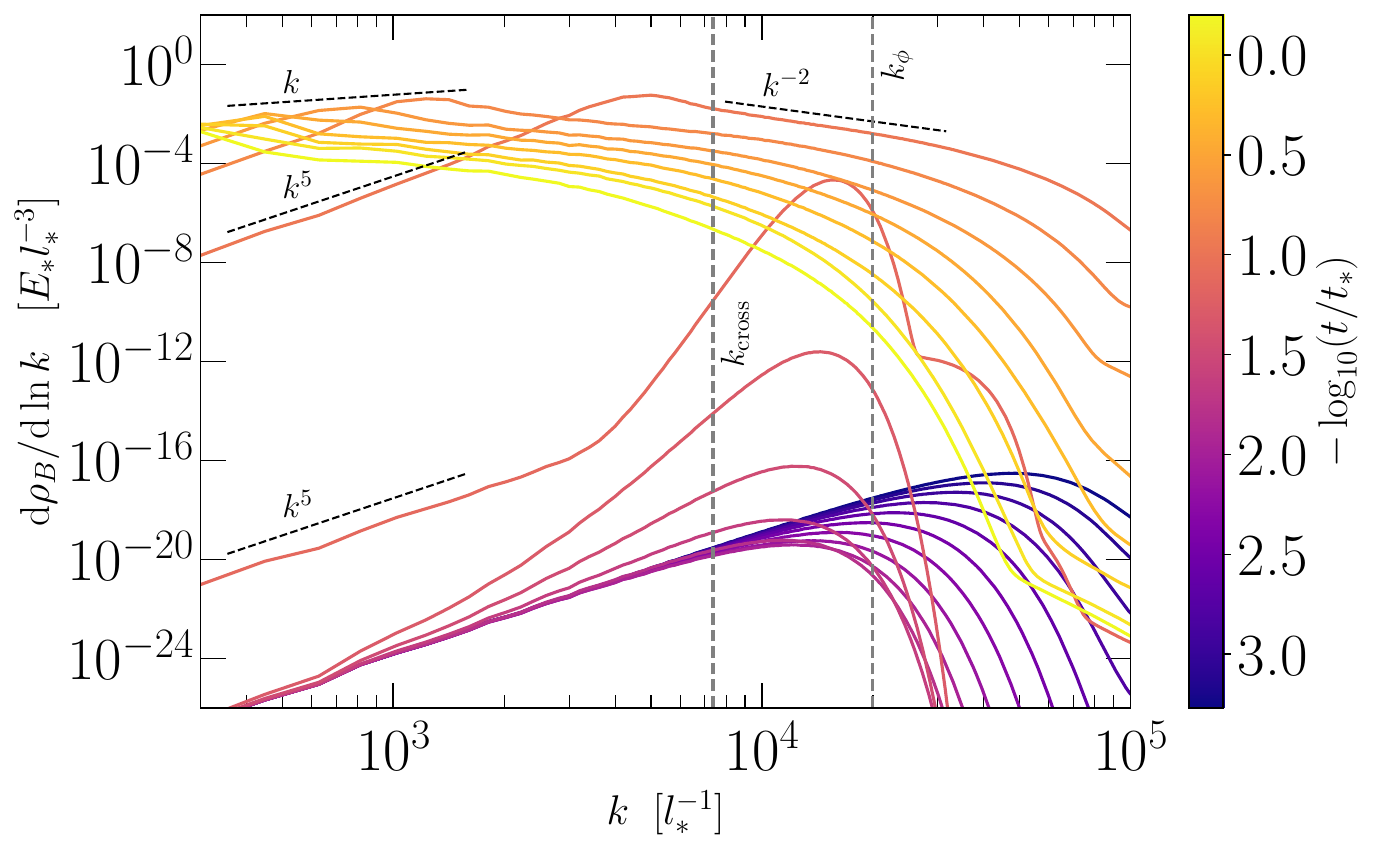}
    \label{fig:spectrum}
    \caption{
    Evolution of the magnetic spectra.  We plot the comoving magnetic energy density $\dd \rho_B / \dd \ln k$ in units of $E_\ast l_\ast^{-3}$ as a function of the comoving wavenumber $k$ in units of $l_\ast^{-1}$ for several values of the conformal time $t$.  Vertical dotted lines represent the inverse length scales in \Eq{eq:wavenumbers}.
    }
\end{figure}

\Fref{fig:spectrum} shows the spectrum of magnetic energy $\dd \rho_B / \dd \ln k$ at several times while the CPI is developing.  
The initial white-noise spectrum \eqref{eq:initial_conditions} corresponds to $\dd\rho_B / \dd \ln k \propto k^5$.  
At early times the magnetic diffusivity suppresses the spectrum at high wavenumber $k$.  
After the time reaches $t \approx t_\mathrm{cross}$, the CPI begins to develop at the scale with wavenumber $k = k_\mathrm{cross} \approx 7 \times 10^{3} \, l_\ast^{-1}$, and the magnetic field grows.  
As the chiral chemical potential continues to increase, the scale of the CPI evolves to larger $k$, and eventually reaches $k = k_\phi \approx 2 \times 10^{4} \, l_\ast^{-1}$ when the source is largest.  
Subsequently, the source decreases rapidly, and the CPI is inactive.  
The magnetic field continues to freely decay.
When the  magnetic field becomes fully helical, the inverse cascade proceeds and the peak of the spectrum evolves toward smaller $k$ while tracing a power-law envelope $\propto k^1$ \cite{Brandenburg:2017neh}.

\begin{figure}[t]
    \centering
    \includegraphics[width=0.95\linewidth]{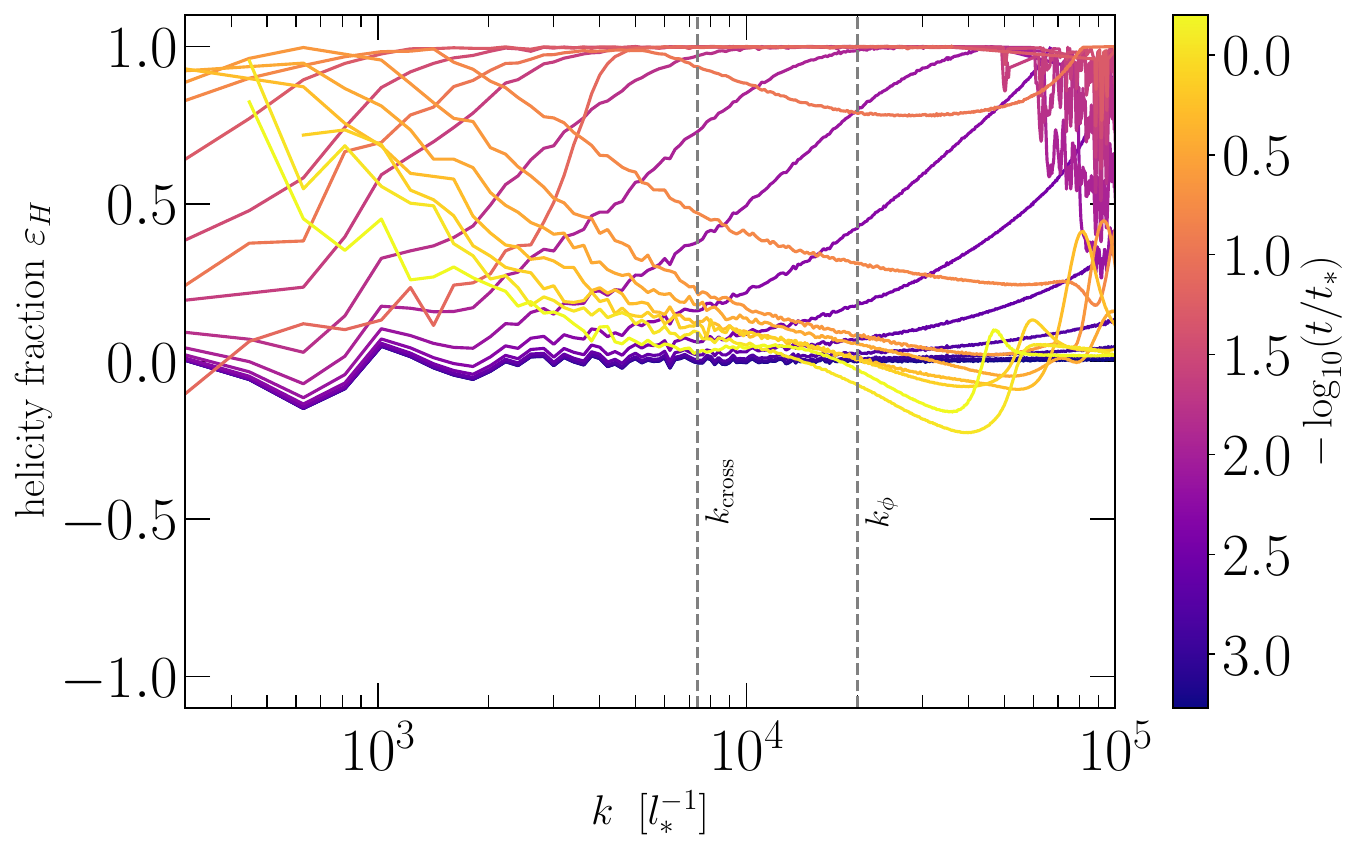}
    \caption{\label{fig:helicity_fraction}
    Helicity fraction.  We plot the comoving magnetic helicity density $\dd h_M / \dd \ln k$ normalized to $\tfrac{2}{k} \dd \rho_B / \dd \ln k$, which is the dimensionless helicity fraction $\varepsilon_H (k,t)$.  The realizability condition requires this ratio to remain $-1 \leq \varepsilon_H \leq +1$, where a maximally-helical field saturates one of the inequalities. Vertical dotted lines represent the length scales in \Eq{eq:wavenumbers}.  
    }
\end{figure}

\Fref{fig:helicity_fraction} shows the helicity fraction of the magnetic field. 
At early times, the helicity of the white noise initial condition is random and small.  
As the high-$k$ modes of the magnetic field decay due to magnetic diffusion, the helicity fraction $\varepsilon_H$ grows, since the denominator decreases more quickly than the numerator. 
The fraction saturates to $\varepsilon_H \approx 1$ corresponding to maximal helicity.  
In fact, $\varepsilon_H$ becomes slightly larger than $1$ as a result of small numerical errors, and the singular nature of $\varepsilon_H \sim 0/0$ when the field is exponentially weak.  
At later times $t > t_\mathrm{cross}$, the CPI grows the seed field, and a maximally helical field is reached for modes with $k \approx k_\mathrm{cross}$.  
As the peak of the spectrum evolves toward smaller $k$ during the inverse cascade, the maximal helicity fraction $\varepsilon_H \approx 1$ also shifts to smaller $k$.  

\begin{figure}[t]
    \centering
    \includegraphics[width=0.95\linewidth]{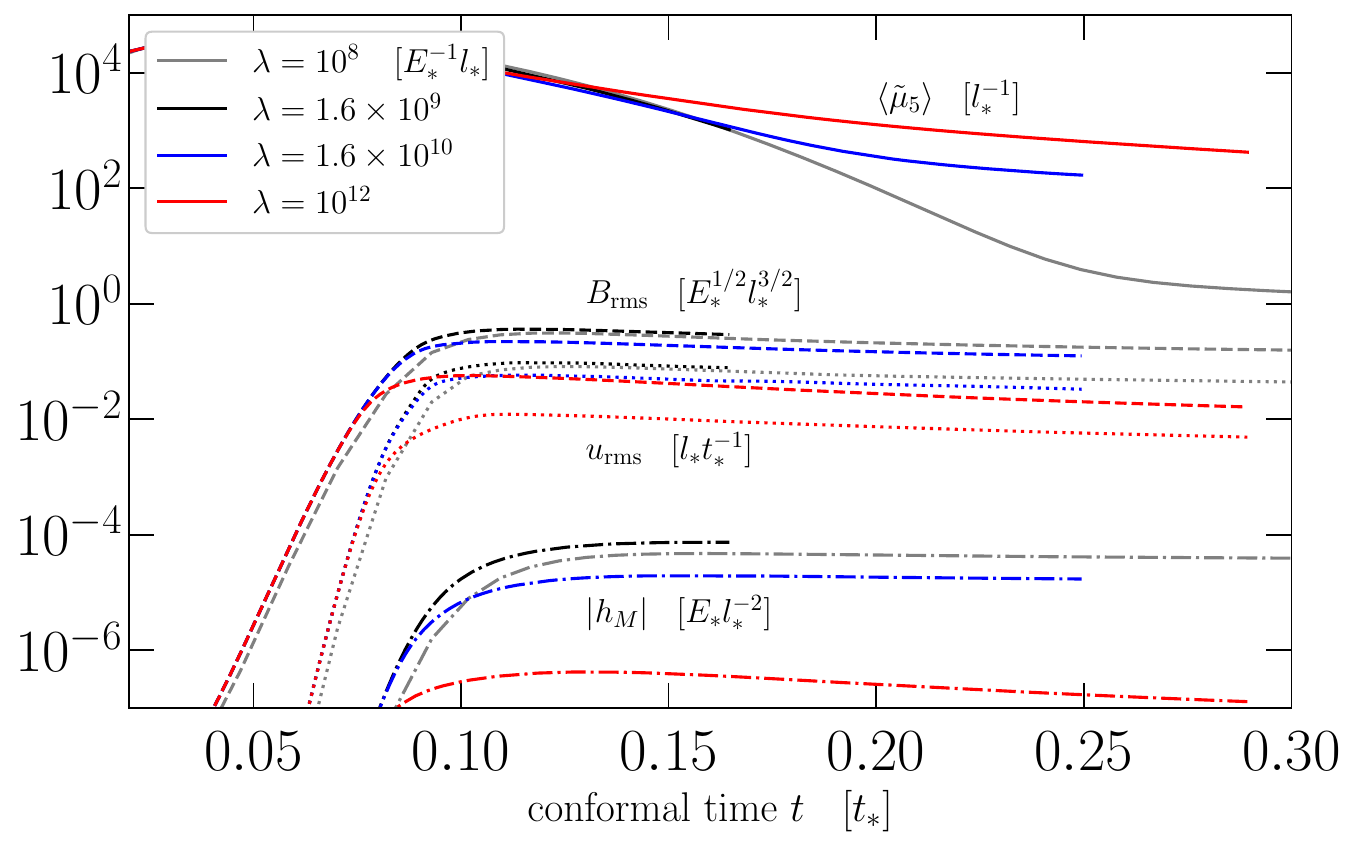}
    \caption{\label{fig:brms_allruns}
    Magnetic (dashed), helicity (dashed-dotted), rms velocity of the fluid (dotted), and chiral chemical potential (solid) fields volume averaged evolution for several values of $\lambda$.  This figure shows late time evolution as \fref{fig:magnetic}, but for each of the parameter sets in \tref{tab:runs}.
    }
\end{figure}

All of the results presented thus far in this section were for parameter set $\mathbf{A}^\ast$ from \tref{tab:runs}.  
In order to get a sense of how varying parameters impacts the predicted magnetic field, we present \fref{fig:brms_allruns}, which shows the magnetic field evolution for several different values of the chiral feedback parameter $\lambda$, corresponding to runs ${\bf B}$, ${\bf C}$, and ${\bf D}$.
Increasing $\lambda$ leads to a stronger back reaction on the evolution of $\tilde{\mu}_5$ when the CPI develops and $\BB$ grows; see \Eq{eq:dmudt}.  
As a result, the final magnetic field is smaller.
Consequently, $u_{\rm rms}$ and $h_M$ are also smaller, while $\tilde \mu_5$ is larger, as inferred from \Eq{eq:mu5_equilib_2}. 

\section{Summary and conclusion}
\label{sec:summary}

Since chirality-violating interactions come into equilibrium when the temperature of the primordial plasma drops below $80 \TeV$, any preexisting chiral asymmetry will be erased.  
However, a source for chirality counteracts its washout, allowing for a nonzero chiral asymmetry even below $80 \TeV$.  
The presence of a chiral asymmetry opens the possibility of helical magnetic field generation via the chiral plasma instability. 
In this work, we have studied the chiral plasma instability in the presence of such a source, together with nonzero chiral erasure rate.

We model the system as a fluid, whose constituents carry both electric and chiral charge, which is therefore coupled to electromagnetism.  
We study the dynamics of this system with the equations of \cMHD{} extended to include a source term.  
We assume a particular functional form for the source's time dependence, which we motivate by considering the out-of-equilibrium decay of a metastable scalar field.  
When selecting parameters, it is important to ensure that the time scale for the CPI is short compared to the lifetime of this scalar field, since otherwise the source will deactivate before the instability has time to develop. 
We solve the equations of \cMHD{} using analytical approximations and numerical methods, the latter implemented in the open-source {\sc Pencil Code}.  

Our calculations reveal that a source of chirality is sufficient to induce the growth of a helical magnetic field via the CPI even at plasma temperatures below $80 \TeV$ where the chirality-washout reactions are in equilibrium.  
We track the evolution of the chiral chemical potential and magnetic helicity in our analytical calculations, which provide an understanding of the general time dependence, and in our numerical calculations, which validate the analytical approximations. 
The resultant magnetic helicity is predicted in two ways.  
The relation in \Eq{eq:hel_prod_source} is the approximate comoving magnetic helicity density that we expect to result from the CPI in the presence of both a source and washout channel for chirality.  
We validate this approximate relation using direct numerical simulation of the \cMHD{} field equations.  
For the parameter values corresponding to $\mathbf{A}^\ast$ in \tref{tab:runs}, our analytical approximation in \Eq{eq:hel_prod_source} gives $h_M \approx (1.6 \times 10^{-13} \G)^2 \Mpc \, a_0^3$, and our numerical calculation gives $h_M \approx (2.4 \times 10^{-13} \G)^2 \Mpc \, a_0^3$ in \Eq{eq:hM_numerical}.  
The excellent agreement gives us confidence in our analytical approximations.  
\Eq{eq:hel_prod_source} is one of the main results of our paper.  

We conclude that when a source for chirality is present, the chiral charge erasure can be avoided.  
This allows the chiral plasma instability to develop even after the chirality-erasing reactions are in thermal equilibrium at plasma temperatures below $80 \TeV$.  
We expect that a sourced CPI may play an important role in magnetic field generation during the cosmological electroweak or QCD phase transition.  
However the parameters that we have explored in our numerical work were chosen to facilitate numerical calculations.  
If we instead use the Standard Model fiducial values of the parameters $\lambda$ and $\Gamma_5$, see \Eq{eq:lambda} and below \Eq{eq:mu5_convert}, then \Eq{eq:hel_prod_source} evaluates to 
\begin{align}
    h_M & \approx \bigl( 3.8 \times 10^{-21} \G \bigr)^2 \Mpc \, a_0^3
    \\ & \quad \times 
    \biggl( \frac{k_B T_{\mathrm{\phys},\ast}}{100 \GeV} \biggr) 
    \biggl( \frac{\epsilon \beta \Omega_\phi}{10^{-5}} \biggr) 
    \biggl( \frac{m_\phi c^2}{100 \GeV} \biggr)^{-1} 
    \nonumber \\ & \quad \times 
    \biggl( \frac{t_\phi}{0.05 \, t_\ast} \biggr)^{-3}
    \biggl( \frac{\Gamma_5}{\Gamma_{5,\star}} \biggr)^{-2} 
    \biggl( \frac{\lambda}{\lambda_\star} \biggr)^{-1} 
    \nonumber
    \;,
\end{align}
where we have also made reasonable choices for the parameters associated with the source, namely $\epsilon$, $\beta$, $\Omega_\phi$, $m_\phi$, and $t_\phi$.  
This level of magnetic helicity would be inadequate to explain blazar observations that provide evidence for a nonzero intergalactic magnetic field.  
Nevertheless, even such a weak field could be strong enough to seed the galactic magnetic fields.  

\bigskip

{\bf Data availability}---The source code used for the
simulations of this study, the {\sc Pencil Code},
is freely available from Ref.~\cite{PC}.
The simulation setups and the corresponding data
are freely available from Ref.~\cite{DATA}.
The calculations and routines
to reproduce the results of this work
will be publicly
available as part of the public Python package {\sc CosmoGW} in its next release \cite{COSMOGW}.

\bigskip
{\bf Acknowledgements} --- 
We thank Balin Armstrong for discussion and comments on the draft. The support through the NASA Astrophysics Theory (ATP) Program award 80NSSC22K0825, the National Science Foundation (NSF) Astronomy and Astrophysics Research Grants (AAG) awards AST2307698 and AST2408411, the  Swedish  Research Council, grant 2019-04234, the Shota Rustaveli GNSF (grant FR24-2606), and the Swiss National Science Foundation
(SNSF Ambizione grant \href{https://data.snf.ch/grants/grant/208807}{208807})
is gratefully acknowledged. We acknowledge the allocation of computing resources provided by the Swedish National Allocations Committee at the Center for Parallel Computers at the Royal Institute of Technology in Stockholm.
AB, MG, and ARP acknowledge the hospitality of CERN during the Pencil Code school
and user meeting,
where part of this work was conducted.
AL and ARP are grateful for the support provided by ICTS during the program
``Hearing beyond the standard model with cosmic sources of Gravitational Waves.''

\bibliographystyle{JHEP}
\bibliography{paper}

@article{Schober:2017cdw,
    author = {Schober, Jennifer and Rogachevskii, Igor and Brandenburg, Axel and Boyarsky, Alexey and Fr{\"o}hlich, J{\"u}rg and Ruchayskiy, Oleg and Kleeorin, Nathan},
    title = "{Laminar and turbulent dynamos in chiral magnetohydrodynamics. II. Simulations}",
    eprint = "1711.09733",
    archivePrefix = "arXiv",
    primaryClass = "physics.flu-dyn",
    reportNumber = "NORDITA-2017-118",
    doi = "10.3847/1538-4357/aaba75",
    journal = "Astrophys. J.",
    volume = "858",
    number = "2",
    pages = "124",
    year = "2018"
}

@article{Schober:2018wlo,
    author = "Schober, J. and Brandenburg, A. and Rogachevskii, I.",
    title = "{Chiral fermion asymmetry in high-energy plasma simulations}",
    eprint = "1808.06624",
    archivePrefix = "arXiv",
    primaryClass = "physics.plasm-ph",
    reportNumber = "NORDITA-2018-076",
    doi = "10.1080/03091929.2019.1591393",
    journal = "Geophys. Astrophys. Fluid Dynamics",
    volume = "114",
    number = "1-2",
    pages = "106--129",
    year = "2020"
}

@article{Schober:2021iws,
    author = "Schober, Jennifer and Rogachevskii, Igor and Brandenburg, Axel",
    title = "{Dynamo instabilities in plasmas with inhomogeneous chiral chemical potential}",
    eprint = "2107.13028",
    archivePrefix = "arXiv",
    primaryClass = "physics.plasm-ph",
    reportNumber = "NORDITA-2021-067",
    doi = "10.1103/PhysRevD.105.043507",
    journal = "Phys. Rev. D",
    volume = "105",
    number = "4",
    pages = "043507",
    year = "2022"
}

@article{Schober:2023zxl,
    author = "Schober, Jennifer and Rogachevskii, Igor and Brandenburg, Axel",
    title = "{Chiral Anomaly and Dynamos from Inhomogeneous Chemical Potential Fluctuations}",
    eprint = "2307.15118",
    archivePrefix = "arXiv",
    primaryClass = "physics.plasm-ph",
    reportNumber = "NORDITA-2023-041",
    doi = "10.1103/PhysRevLett.132.065101",
    journal = "Phys. Rev. Lett.",
    volume = "132",
    number = "6",
    pages = "065101",
    year = "2024"
}

@article{Schober:2024vtv,
    author = "Schober, Jennifer and Rogachevskii, Igor and Brandenburg, Axel",
    title = "{Efficiency of dynamos from an autonomous generation of chiral asymmetry}",
    eprint = "2404.07845",
    archivePrefix = "arXiv",
    primaryClass = "physics.plasm-ph",
    reportNumber = "NORDITA-2024-008",
    doi = "10.1103/PhysRevD.110.043515",
    journal = "Phys. Rev. D",
    volume = "110",
    number = "4",
    pages = "043515",
    year = "2024"
}

@article{Hattori:2017usa,
    author = "Hattori, Koichi and Hirono, Yuji and Yee, Ho-Ung and Yin, Yi",
    title = "{MagnetoHydrodynamics with chiral anomaly: phases of collective excitations and instabilities}",
    eprint = "1711.08450",
    archivePrefix = "arXiv",
    primaryClass = "hep-th",
    reportNumber = "MIT-CTP/4958, MIT-CTP-4958",
    doi = "10.1103/PhysRevD.100.065023",
    journal = "Phys. Rev. D",
    volume = "100",
    number = "6",
    pages = "065023",
    year = "2019"
}

@article{Giovannini:2013oga,
    author = "Giovannini, Massimo",
    title = "{Anomalous Magnetohydrodynamics}",
    eprint = "1307.2454",
    archivePrefix = "arXiv",
    primaryClass = "hep-th",
    reportNumber = "CERN-PH-TH-2013-152",
    doi = "10.1103/PhysRevD.88.063536",
    journal = "Phys. Rev. D",
    volume = "88",
    pages = "063536",
    year = "2013"
}

@article{Yamamoto:2016xtu,
    author = "Yamamoto, Naoki",
    title = "{Scaling laws in chiral hydrodynamic turbulence}",
    eprint = "1603.08864",
    archivePrefix = "arXiv",
    primaryClass = "hep-th",
    doi = "10.1103/PhysRevD.93.125016",
    journal = "Phys. Rev. D",
    volume = "93",
    number = "12",
    pages = "125016",
    year = "2016"
}

@article{Boyarsky:2011uy,
    author = "Boyarsky, Alexey and Fr{\"o}hlich, Jurg and Ruchayskiy, Oleg",
    title = "{Self-consistent evolution of magnetic fields and chiral asymmetry in the early Universe}",
    eprint = "1109.3350",
    archivePrefix = "arXiv",
    primaryClass = "astro-ph.CO",
    reportNumber = "CERN-PH-TH-2011-231",
    doi = "10.1103/PhysRevLett.108.031301",
    journal = "Phys. Rev. Lett.",
    volume = "108",
    pages = "031301",
    year = "2012"
}

@article{Brandenburg:2023aco,
       author = {{Brandenburg}, Axel and {Kamada}, Kohei and {Mukaida}, Kyohei and {Schmitz}, Kai and {Schober}, Jennifer},
        title = "{Chiral magnetohydrodynamics with zero total chirality}",
      journal = {\prd},
         year = 2023,
        month = sep,
       volume = {108},
       number = {6},
          eid = {063529},
        pages = {063529},
          doi = {10.1103/PhysRevD.108.063529},
archivePrefix = {arXiv},
       eprint = {2304.06612},
 primaryClass = {hep-ph},
       adsurl = {https://ui.adsabs.harvard.edu/abs/2023PhRvD.108f3529B}
}

@article{Schober:2020ogz,
    author = "Schober, Jennifer and Schober, Jennifer and Fujita, Tomohiro and Fujita, Tomohiro and Durrer, Ruth and Durrer, Ruth",
    title = "{Generation of chiral asymmetry via helical magnetic fields}",
    eprint = "2002.09501",
    archivePrefix = "arXiv",
    primaryClass = "physics.plasm-ph",
    doi = "10.1103/PhysRevD.101.103028",
    journal = "Phys. Rev. D",
    volume = "101",
    number = "10",
    pages = "103028",
    year = "2020",
    note = "[Erratum: Phys.Rev.D 105, 069901 (2022)]"
}

@article{Schober:2021yav,
    author = "Schober, Jennifer and Rogachevskii, Igor and Brandenburg, Axel",
    title = "{Production of a Chiral Magnetic Anomaly with Emerging Turbulence and Mean-Field Dynamo Action}",
    eprint = "2107.12945",
    archivePrefix = "arXiv",
    primaryClass = "physics.plasm-ph",
    reportNumber = "NORDITA-2021-066",
    doi = "10.1103/PhysRevLett.128.065002",
    journal = "Phys. Rev. Lett.",
    volume = "128",
    number = "6",
    pages = "065002",
    year = "2022"
}

@article{Akamatsu:2013pjd,
    author = "Akamatsu, Yukinao and Yamamoto, Naoki",
    title = "{Chiral Plasma Instabilities}",
    eprint = "1302.2125",
    archivePrefix = "arXiv",
    primaryClass = "nucl-th",
    reportNumber = "YITP-13-9",
    doi = "10.1103/PhysRevLett.111.052002",
    journal = "Phys. Rev. Lett.",
    volume = "111",
    pages = "052002",
    year = "2013"
}

@article{Kamada:2022nyt,
    author = "Kamada, Kohei and Yamamoto, Naoki and Yang, Di-Lun",
    title = "{Chiral effects in astrophysics and cosmology}",
    eprint = "2207.09184",
    archivePrefix = "arXiv",
    primaryClass = "astro-ph.CO",
    reportNumber = "RESCEU-12/22",
    doi = "10.1016/j.ppnp.2022.104016",
    journal = "Prog. Part. Nucl. Phys.",
    volume = "129",
    pages = "104016",
    year = "2023"
}

@article{Boyarsky:2012ex,
    author = "Boyarsky, Alexey and Ruchayskiy, Oleg and Shaposhnikov, Mikhail",
    title = "{Long-range magnetic fields in the ground state of the Standard Model plasma}",
    eprint = "1204.3604",
    archivePrefix = "arXiv",
    primaryClass = "hep-ph",
    reportNumber = "CERN-PH-TH-2012-112",
    doi = "10.1103/PhysRevLett.109.111602",
    journal = "Phys. Rev. Lett.",
    volume = "109",
    pages = "111602",
    year = "2012"
}

@article{Sakharov:1967dj,
    author = "Sakharov, A. D.",
    title = "{Violation of CP Invariance, C asymmetry, and baryon asymmetry of the universe}",
    doi = "10.1070/PU1991v034n05ABEH002497",
    journal = "Pisma Zh. Eksp. Teor. Fiz.",
    volume = "5",
    pages = "32--35",
    year = "1967"
}

@article{Kharzeev:2013ffa,
    author = "Kharzeev, Dmitri E.",
    title = "{The Chiral Magnetic Effect and Anomaly-Induced Transport}",
    eprint = "1312.3348",
    archivePrefix = "arXiv",
    primaryClass = "hep-ph",
    doi = "10.1016/j.ppnp.2014.01.002",
    journal = "Prog. Part. Nucl. Phys.",
    volume = "75",
    pages = "133--151",
    year = "2014"
}

@article{Vilenkin:1980fu,
    author = "Vilenkin, A.",
    title = "{Equilibrium Parity Violating Current in a Magnetic Field}",
    doi = "10.1103/PhysRevD.22.3080",
    journal = "Phys. Rev. D",
    volume = "22",
    pages = "3080--3084",
    year = "1980"
}

@misc{PC,
	title = {The Pencil Code.  DOI:10.5281/zenodo.2315093.  \href{https://github.com/pencil-code}{https://github.com/pencil-code}}}

@article{PencilCode:2020eyn,
    author = "Brandenburg, A. and others",
    collaboration = "Pencil Code Collaboration",
    title = "{The Pencil Code, a modular MPI code for partial differential equations and particles: multipurpose and multiuser-maintained}",
    eprint = "2009.08231",
    archivePrefix = "arXiv",
    primaryClass = "astro-ph.IM",
    reportNumber = "NORDITA-2020-087",
    doi = "10.21105/joss.02807",
    journal = "J. Open Source Softw.",
    volume = "6",
    number = "58",
    pages = "2807",
    year = "2021"
}

@article{Fujita:2016igl,
	archiveprefix = {arXiv},
	author = {Fujita, Tomohiro and Kamada, Kohei},
	doi = {10.1103/PhysRevD.93.083520},
	eprint = {1602.02109},
	journal = {Phys. Rev. D},
	number = {8},
	pages = {083520},
	primaryclass = {hep-ph},
	title = {{Large-scale magnetic fields can explain the baryon asymmetry of the Universe}},
	volume = {93},
	year = {2016}}

@article{Kamada:2016cnb,
	archiveprefix = {arXiv},
	author = {Kamada, Kohei and Long, Andrew J.},
	doi = {10.1103/PhysRevD.94.123509},
	eprint = {1610.03074},
	journal = {Phys. Rev. D},
	number = {12},
	pages = {123509},
	primaryclass = {hep-ph},
	title = {{Evolution of the Baryon Asymmetry through the Electroweak Crossover in the Presence of a Helical Magnetic Field}},
	volume = {94},
	year = {2016}}

@article{Kamada:2016eeb,
	archiveprefix = {arXiv},
	author = {Kamada, Kohei and Long, Andrew J.},
	doi = {10.1103/PhysRevD.94.063501},
	eprint = {1606.08891},
	journal = {Phys. Rev. D},
	number = {6},
	pages = {063501},
	primaryclass = {astro-ph.CO},
	title = {{Baryogenesis from decaying magnetic helicity}},
	volume = {94},
	year = {2016}}

@article{Rogachevskii:2017uyc,
    author = {Rogachevskii, Igor and Ruchayskiy, Oleg and Boyarsky, Alexey and Fr{\"o}hlich, J\"urg and Kleeorin, Nathan and Brandenburg, Axel and Schober, Jennifer},
    title = "{Laminar and turbulent dynamos in chiral magnetohydrodynamics-I: Theory}",
    eprint = "1705.00378",
    archivePrefix = "arXiv",
    primaryClass = "physics.plasm-ph",
    reportNumber = "PREPRINT-NORDITA-2017-39",
    doi = "10.3847/1538-4357/aa886b",
    journal = "Astrophys. J.",
    volume = "846",
    number = "2",
    pages = "153",
    year = "2017"
}

@article{Brandenburg:2017rcb,
	archiveprefix = {arXiv},
	author = {Brandenburg, Axel and Schober, Jennifer and Rogachevskii, Igor and Kahniashvili, Tina and Boyarsky, Alexey and Fr{\"o}hlich, Jurg and Ruchayskiy, Oleg and Kleeorin, Nathan},
	doi = {10.3847/2041-8213/aa855d},
	eprint = {1707.03385},
	journal = {Astrophys. J. Lett.},
	number = {2},
	pages = {L21},
	primaryclass = {astro-ph.CO},
	reportnumber = {NORDITA-2017-037},
	title = {{The turbulent chiral-magnetic cascade in the early universe}},
	volume = {845},
	year = {2017}}

@article{Boyarsky:2015faa,
	archiveprefix = {arXiv},
	author = {Boyarsky, Alexey and Fr{\"o}hlich, Jurg and Ruchayskiy, Oleg},
	doi = {10.1103/PhysRevD.92.043004},
	eprint = {1504.04854},
	journal = {Phys. Rev. D},
	pages = {043004},
	primaryclass = {hep-ph},
	title = {{Magnetohydrodynamics of Chiral Relativistic Fluids}},
	volume = {92},
	year = {2015}}

@article{Neronov:2010gir,
	archiveprefix = {arXiv},
	author = {Neronov, A. and Vovk, I.},
	doi = {10.1126/science.1184192},
	eprint = {1006.3504},
	journal = {Science},
	pages = {73--75},
	primaryclass = {astro-ph.HE},
	title = {{Evidence for strong extragalactic magnetic fields from Fermi observations of TeV blazars}},
	volume = {328},
	year = {2010}}

@article{Trodden:1998ym,
	archiveprefix = {arXiv},
	author = {Trodden, Mark},
	doi = {10.1103/RevModPhys.71.1463},
	eprint = {hep-ph/9803479},
	journal = {Rev. Mod. Phys.},
	pages = {1463--1500},
	reportnumber = {CWRU-P6-98},
	title = {{Electroweak baryogenesis}},
	volume = {71},
	year = {1999}}

@article{Brandenburg:2017neh,
	archiveprefix = {arXiv},
	author = {Brandenburg, Axel and Kahniashvili, Tina and Mandal, Sayan and Roper Pol, Alberto and Tevzadze, Alexander G. and Vachaspati, Tanmay},
	doi = {10.1103/PhysRevD.96.123528},
	eprint = {1711.03804},
	journal = {Phys. Rev. D},
	number = {12},
	pages = {123528},
	primaryclass = {astro-ph.CO},
	reportnumber = {NORDITA-2017-116},
	title = {{Evolution of hydromagnetic turbulence from the electroweak phase transition}},
	volume = {96},
	year = {2017}}

@article{Vachaspati:2020blt,
	archiveprefix = {arXiv},
	author = {Vachaspati, Tanmay},
	doi = {10.1088/1361-6633/ac03a9},
	eprint = {2010.10525},
	journal = {Rept. Prog. Phys.},
	number = {7},
	pages = {074901},
	primaryclass = {astro-ph.CO},
	title = {{Progress on cosmological magnetic fields}},
	volume = {84},
	year = {2021}}

@article{Brandenburg:2021aln,
	archiveprefix = {arXiv},
	author = {Brandenburg, Axel and He, Yutong and Kahniashvili, Tina and Rheinhardt, Matthias and Schober, Jennifer},
	doi = {10.3847/1538-4357/abe4d7},
	eprint = {2101.08178},
	journal = {Astrophys. J.},
	number = {2},
	pages = {110},
	primaryclass = {astro-ph.CO},
	reportnumber = {NORDITA-2021-005},
	title = {{Relic gravitational waves from the chiral magnetic effect}},
	volume = {911},
	year = {2021}}

@article{Brandenburg:1996fc,
	archiveprefix = {arXiv},
	author = {Brandenburg, Axel and Enqvist, Kari and Olesen, Poul},
	doi = {10.1103/PhysRevD.54.1291},
	eprint = {astro-ph/9602031},
	journal = {Phys. Rev. D},
	pages = {1291--1300},
	reportnumber = {NORDITA-96-6-A},
	title = {{Large scale magnetic fields from hydromagnetic turbulence in the very early universe}},
	volume = {54},
	year = {1996}}

@article{Bodeker:2019ajh,
    author = {B\"odeker, Dietrich and Schr\"oder, Dennis},
    title = "{Equilibration of right-handed electrons}",
    eprint = "1902.07220",
    archivePrefix = "arXiv",
    primaryClass = "hep-ph",
    doi = "10.1088/1475-7516/2019/05/010",
    journal = "JCAP",
    volume = "05",
    pages = "010",
    year = "2019"
}

@article{Durrer:2013pga,
    author = "Durrer, Ruth and Neronov, Andrii",
    title = "{Cosmological Magnetic Fields: Their Generation, Evolution and Observation}",
    eprint = "1303.7121",
    archivePrefix = "arXiv",
    primaryClass = "astro-ph.CO",
    doi = "10.1007/s00159-013-0062-7",
    journal = "Astron. Astrophys. Rev.",
    volume = "21",
    pages = "62",
    year = "2013"
}

@article{Jedamzik:2020krr,
    author = "Jedamzik, Karsten and Pogosian, Levon",
    title = "{Relieving the Hubble tension with primordial magnetic fields}",
    eprint = "2004.09487",
    archivePrefix = "arXiv",
    primaryClass = "astro-ph.CO",
    doi = "10.1103/PhysRevLett.125.181302",
    journal = "Phys. Rev. Lett.",
    volume = "125",
    number = "18",
    pages = "181302",
    year = "2020"
}

@article{Joyce:1997uy,
	archiveprefix = {arXiv},
	author = {Joyce, M. and Shaposhnikov, Mikhail E.},
	doi = {10.1103/PhysRevLett.79.1193},
	eprint = {astro-ph/9703005},
	journal = {Phys.Rev.Lett.},
	pages = {1193-1196},
	primaryclass = {astro-ph},
	reportnumber = {CERN-TH-97-31},
	slaccitation = {%%CITATION = ASTRO-PH/9703005;%%},
	title = {{Primordial magnetic fields, right-handed electrons, and the Abelian anomaly}},
	volume = {79},
	year = {1997}}

@article{Campbell:1990fa,
	author = {Campbell, Bruce A. and Davidson, Sacha and Ellis, John R. and Olive, Keith A.},
	doi = {10.1016/0370-2693(91)91795-W},
	journal = {Phys.Lett.},
	pages = {484-490},
	reportnumber = {CERN-TH-5926-90, ALBERTA-THY-32-90, LAPP-TH-321-90, UMN-TH-915-90},
	slaccitation = {%%CITATION = PHLTA,B256,484;%%},
	title = {{Cosmological baryon asymmetry constraints on extensions of the standard model}},
	volume = {B256},
	year = {1991}}

@article{Arnold:2000dr,
    author = "Arnold, Peter Brockway and Moore, Guy D. and Yaffe, Laurence G.",
    title = "{Transport coefficients in high temperature gauge theories. 1. Leading log results}",
    eprint = "hep-ph/0010177",
    archivePrefix = "arXiv",
    reportNumber = "UW-PT-00-15",
    doi = "10.1088/1126-6708/2000/11/001",
    journal = "JHEP",
    volume = "11",
    pages = "001",
    year = "2000"
}

@misc{DATA,
      title = {Datasets for ``{P}rimordial magnetic field from chiral plasma instability with sourcing'' v.2025.08.12, \href{https://doi.org/10.5281/zenodo.17852669}{doi.org/10.5281/zenodo.17852669}; see \url{http://norlx65.nordita.org/~brandenb/proj/Source-CME/}}
}

@misc{COSMOGW,
    title        = {Public code for reproducing results of this work},
    note         = {GitHub repository},
    howpublished = {\url{https://github.com/cosmoGW/cosmoGW}}
}

@article{RoperPol:2025lgc,
    author = "Roper Pol, Alberto and Midiri, Antonino Salvino",
    title = "{Relativistic magnetohydrodynamics in the early Universe}",
    eprint = "2501.05732",
    archivePrefix = "arXiv",
    primaryClass = "gr-qc",
    month = "1",
    year = "2025"
}

@article{Tashiro:2012mf,
    author = "Tashiro, Hiroyuki and Vachaspati, Tanmay and Vilenkin, Alexander",
    title = "{Chiral Effects and Cosmic Magnetic Fields}",
    eprint = "1206.5549",
    archivePrefix = "arXiv",
    primaryClass = "astro-ph.CO",
    doi = "10.1103/PhysRevD.86.105033",
    journal = "Phys. Rev. D",
    volume = "86",
    pages = "105033",
    year = "2012"
}

@article{Brandenburg:2023imm,
    author = "Brandenburg, Axel and Clarke, Emma and Kahniashvili, Tina and Long, Andrew J. and Sun, Guotong",
    title = "{Relic gravitational waves from the chiral plasma instability in the standard cosmological model}",
    eprint = "2307.09385",
    archivePrefix = "arXiv",
    primaryClass = "astro-ph.CO",
    reportNumber = "NORDITA-2023-034",
    doi = "10.1103/PhysRevD.109.043534",
    journal = "Phys. Rev. D",
    volume = "109",
    number = "4",
    pages = "043534",
    year = "2024"
}

@book{Peskin:1995,
	address = {{B}oulder, {C}olorado},
	author = {Peskin, Michael E. and Schroeder, Daniel V.},
	publisher = {{W}estview {P}ress},
	title = {{A}n {I}ntroduction to {Q}uantum {F}ield {T}heory},
	year = {1995}}

@article{Long:2016uez,
    author = "Long, Andrew J. and Sabancilar, Eray",
    title = "{Chiral Charge Erasure via Thermal Fluctuations of Magnetic Helicity}",
    eprint = "1601.03777",
    archivePrefix = "arXiv",
    primaryClass = "hep-th",
    doi = "10.1088/1475-7516/2016/05/029",
    journal = "JCAP",
    volume = "05",
    pages = "029",
    year = "2016"
}

@article{Pilaftsis:1997jf,
    author = "Pilaftsis, Apostolos",
    title = "{CP violation and baryogenesis due to heavy Majorana neutrinos}",
    eprint = "hep-ph/9707235",
    archivePrefix = "arXiv",
    reportNumber = "MPI-PHT-97-30",
    doi = "10.1103/PhysRevD.56.5431",
    journal = "Phys. Rev. D",
    volume = "56",
    pages = "5431--5451",
    year = "1997"
}

@article{Shaposhnikov:1987tw,
    author = "Shaposhnikov, M. E.",
    title = "{Baryon Asymmetry of the Universe in Standard Electroweak Theory}",
    doi = "10.1016/0550-3213(87)90127-1",
    journal = "Nucl. Phys. B",
    volume = "287",
    pages = "757--775",
    year = "1987"
}

@article{Jedamzik:2018itu,
    author = "Jedamzik, Karsten and Saveliev, Andrey",
    title = "{Stringent Limit on Primordial Magnetic Fields from the Cosmic Microwave Background Radiation}",
    eprint = "1804.06115",
    archivePrefix = "arXiv",
    primaryClass = "astro-ph.CO",
    doi = "10.1103/PhysRevLett.123.021301",
    journal = "Phys. Rev. Lett.",
    volume = "123",
    number = "2",
    pages = "021301",
    year = "2019"
}

@article{Subramanian:2019jyd,
    author = "Subramanian, Kandaswamy",
    title = "{From primordial seed magnetic fields to the galactic dynamo}",
    eprint = "1903.03744",
    archivePrefix = "arXiv",
    primaryClass = "astro-ph.CO",
    doi = "10.3390/galaxies7020047",
    journal = "Galaxies",
    volume = "7",
    number = "2",
    pages = "47",
    year = "2019"
}

@article{Jedamzik:2011cu,
    author = "Jedamzik, Karsten and Abel, Tom",
    title = "{Weak Primordial Magnetic Fields and Anisotropies in the Cosmic Microwave Background Radiation}",
    eprint = "1108.2517",
    archivePrefix = "arXiv",
    primaryClass = "astro-ph.CO",
    month = "8",
    year = "2011"
}

@article{Galli:2021mxk,
    author = "Galli, Silvia and Pogosian, Levon and Jedamzik, Karsten and Balkenhol, Lennart",
    title = "{Consistency of Planck, ACT, and SPT constraints on magnetically assisted recombination and forecasts for future experiments}",
    eprint = "2109.03816",
    archivePrefix = "arXiv",
    primaryClass = "astro-ph.CO",
    doi = "10.1103/PhysRevD.105.023513",
    journal = "Phys. Rev. D",
    volume = "105",
    number = "2",
    pages = "023513",
    year = "2022"
}

@article{Mirpoorian:2024fka,
    author = "Mirpoorian, Seyed Hamidreza and Jedamzik, Karsten and Pogosian, Levon",
    title = "{Modified recombination and the Hubble tension}",
    eprint = "2411.16678",
    archivePrefix = "arXiv",
    primaryClass = "astro-ph.CO",
    doi = "10.1103/PhysRevD.111.083519",
    journal = "Phys. Rev. D",
    volume = "111",
    number = "8",
    pages = "083519",
    year = "2025"
}

@article{Jedamzik:2025cax,
    author = "Jedamzik, Karsten and Pogosian, Levon and Abel, Tom",
    title = "{Hints of Primordial Magnetic Fields at Recombination and Implications for the Hubble Tension}",
    eprint = "2503.09599",
    archivePrefix = "arXiv",
    primaryClass = "astro-ph.CO",
    reportNumber = "SCG-2025-01",
    month = "3",
    year = "2025"
}

@article{Fukuda:2025nmc,
    author = "Fukuda, Hajime and Hamada, Yuta and Kamada, Kohei and Mukaida, Kyohei and Uchida, Fumio",
    title = "{Magnetic Helicity, Magnetic Monopoles, and Higgs Winding}",
    eprint = "2509.25734",
    archivePrefix = "arXiv",
    primaryClass = "hep-ph",
    reportNumber = "KEK-TH-2761, IPMU25-0048, KEK-TH-2761, IPMU25-0048",
    month = "9",
    year = "2025"
}

@article{Hamada:2025cwu,
    author = "Hamada, Yuta and Mukaida, Kyohei and Uchida, Fumio",
    title = "{Symmetries of Hot SM, Magnetic Flux {\&} Baryogenesis from Helicity Decay}",
    eprint = "2507.01576",
    archivePrefix = "arXiv",
    primaryClass = "hep-ph",
    reportNumber = "KEK-TH-2737, IPMU25-0035",
    month = "7",
    year = "2025"
}

@article{Pavlovic:2018jiz,
    author = {Pavlovi{\'c}, Petar and Sigl, G{\"u}nter},
    title = "{On minimal energy states of chiral MHD turbulence}",
    eprint = "1806.06447",
    archivePrefix = "arXiv",
    primaryClass = "hep-th",
    doi = "10.1088/1475-7516/2019/04/055",
    journal = "JCAP",
    volume = "04",
    pages = "055",
    year = "2019"
}

@article{Pavlovic:2016gac,
    author = {Pavlovi{\'c}, Petar and Leite, Natacha and Sigl, G{\"u}nter},
    title = "{Chiral Magnetohydrodynamic Turbulence}",
    eprint = "1612.07382",
    archivePrefix = "arXiv",
    primaryClass = "astro-ph.CO",
    doi = "10.1103/PhysRevD.96.023504",
    journal = "Phys. Rev. D",
    volume = "96",
    number = "2",
    pages = "023504",
    year = "2017"
}

@article{Sigl:2015xva,
    author = {Sigl, G{\"u}nter and Leite, Natacha},
    title = "{Chiral Magnetic Effect in Protoneutron Stars and Magnetic Field Spectral Evolution}",
    eprint = "1507.04983",
    archivePrefix = "arXiv",
    primaryClass = "astro-ph.HE",
    doi = "10.1088/1475-7516/2016/01/025",
    journal = "JCAP",
    volume = "01",
    pages = "025",
    year = "2016"
}

@inbook{Kharzeev:2023zbo,
    author = "Kharzeev, Dmitri E.",
    editor = "Tanihata, Isao and Toki, Hiroshi and Kajino, Toshitaka",
    title = "{Chiral Magnetic Effect: A Brief Introduction}",
    booktitle = "{Handbook of Nuclear Physics}",
    doi = "10.1007/978-981-15-8818-1_25-1",
    pages = "1--14",
    year = "2023"
}

@article{Fukushima:2008xe,
    author = "Fukushima, Kenji and Kharzeev, Dmitri E. and Warringa, Harmen J.",
    title = "{The Chiral Magnetic Effect}",
    eprint = "0808.3382",
    archivePrefix = "arXiv",
    primaryClass = "hep-ph",
    doi = "10.1103/PhysRevD.78.074033",
    journal = "Phys. Rev. D",
    volume = "78",
    pages = "074033",
    year = "2008"
}

@article{Kharzeev:2015znc,
    author = "Kharzeev, D. E. and Liao, J. and Voloshin, S. A. and Wang, G.",
    title = "{Chiral magnetic and vortical effects in high-energy nuclear collisions{\textemdash}A status report}",
    eprint = "1511.04050",
    archivePrefix = "arXiv",
    primaryClass = "hep-ph",
    doi = "10.1016/j.ppnp.2016.01.001",
    journal = "Prog. Part. Nucl. Phys.",
    volume = "88",
    pages = "1--28",
    year = "2016"
}

@article{Liao:2014ava,
    author = "Liao, Jinfeng",
    title = "{Anomalous transport effects and possible environmental symmetry {\textquoteleft}violation{\textquoteright} in heavy-ion collisions}",
    eprint = "1401.2500",
    archivePrefix = "arXiv",
    primaryClass = "hep-ph",
    doi = "10.1007/s12043-015-0984-x",
    journal = "Pramana",
    volume = "84",
    number = "5",
    pages = "901--926",
    year = "2015"
}

@article{Akhmedov:1998qx,
    author = "Akhmedov, Evgeny K. and Rubakov, V. A. and Smirnov, A. Yu.",
    title = "{Baryogenesis via neutrino oscillations}",
    eprint = "hep-ph/9803255",
    archivePrefix = "arXiv",
    reportNumber = "IC-98-22, INR-98-14-T",
    doi = "10.1103/PhysRevLett.81.1359",
    journal = "Phys. Rev. Lett.",
    volume = "81",
    pages = "1359--1362",
    year = "1998"
}

@article{Elor:2018twp,
    author = "Elor, Gilly and Escudero, Miguel and Nelson, Ann",
    title = "{Baryogenesis and Dark Matter from $B$ Mesons}",
    eprint = "1810.00880",
    archivePrefix = "arXiv",
    primaryClass = "hep-ph",
    reportNumber = "KCL-18-53, IFIC-18-35",
    doi = "10.1103/PhysRevD.99.035031",
    journal = "Phys. Rev. D",
    volume = "99",
    number = "3",
    pages = "035031",
    year = "2019"
}

@inproceedings{Cline:2006ts,
    author = "Cline, James M.",
    title = "{Baryogenesis}",
    booktitle = "{Les Houches Summer School - Session 86: Particle Physics and Cosmology: The Fabric of Spacetime}",
    eprint = "hep-ph/0609145",
    archivePrefix = "arXiv",
    month = "9",
    year = "2006"
}

@article{Nelson:2019fln,
    author = "Nelson, Ann E. and Xiao, Huangyu",
    title = "{Baryogenesis from B Meson Oscillations}",
    eprint = "1901.08141",
    archivePrefix = "arXiv",
    primaryClass = "hep-ph",
    doi = "10.1103/PhysRevD.100.075002",
    journal = "Phys. Rev. D",
    volume = "100",
    number = "7",
    pages = "075002",
    year = "2019"
}

@article{Klaric:2021cpi,
    author = "Klari{\'c}, Juraj and Shaposhnikov, Mikhail and Timiryasov, Inar",
    title = "{Reconciling resonant leptogenesis and baryogenesis via neutrino oscillations}",
    eprint = "2103.16545",
    archivePrefix = "arXiv",
    primaryClass = "hep-ph",
    doi = "10.1103/PhysRevD.104.055010",
    journal = "Phys. Rev. D",
    volume = "104",
    number = "5",
    pages = "055010",
    year = "2021"
}

@article{MAGIC:2022piy,
    author = "Acciari, V. A. and others",
    collaboration = "MAGIC",
    title = "{A lower bound on intergalactic magnetic fields from time variability of 1ES 0229+200 from MAGIC and Fermi/LAT observations}",
    eprint = "2210.03321",
    archivePrefix = "arXiv",
    primaryClass = "astro-ph.HE",
    doi = "10.1051/0004-6361/202244126",
    journal = "Astron. Astrophys.",
    volume = "670",
    pages = "A145",
    year = "2023"
}

@article{Li:2025yxx,
    author = "Li, Wei and Shou, Qiye and Wang, Fuqiang",
    title = "{Experimental review on the chiral magnetic effect in relativistic heavy ion collisions}",
    eprint = "2511.07358",
    archivePrefix = "arXiv",
    primaryClass = "nucl-ex",
    month = "11",
    year = "2025"
}

\end{document}